\documentclass[preprint,journal]{vgtc}       % preprint (journal style)

\ifpdf%                                % if we use pdflatex
  \pdfoutput=1\relax                   % create PDFs from pdfLaTeX
  \usepackage{graphicx}                % allow us to embed graphics files
  \DeclareGraphicsExtensions{.pdf,.png,.jpg,.jpeg} % for pdflatex we expect .pdf, .png, or .jpg files
\else%                                 % else we use pure latex
  \usepackage{graphicx}                % allow us to embed graphics files
  \DeclareGraphicsExtensions{.eps}     % for pure latex we expect eps files
\fi%

\graphicspath{{figures/}{pictures/}{images/}{./}} % where to search for the images

\usepackage{microtype}                 % use micro-typography (slightly more compact, better to read)
\PassOptionsToPackage{warn}{textcomp}  % to address font issues with \textrightarrow
\usepackage{textcomp}                  % use better special symbols
\usepackage{mathptmx}                  % use matching math font
\usepackage{times}                     % we use Times as the main font
\usepackage{cite}                      % needed to automatically sort the references
\usepackage{tabu}                      % only used for the table example
\usepackage{booktabs}                  % only used for the table example
\usepackage[normalem]{ulem}
\usepackage{amsmath}
\usepackage{amsfonts}
\usepackage{amsfonts}
\usepackage{enumitem}
\usepackage{wrapfig}
\usepackage{pdfpages}
\usepackage{threeparttable,caption,tabularx,booktabs}
\usepackage{xcolor}
\usepackage{algorithm,algcompatible}
\usepackage[noend]{algpseudocode}
\usepackage{amsmath}
\usepackage{forloop}
\newcolumntype{L}{>{\hsize=1.5\hsize\raggedright\arraybackslash}X}
\newcolumntype{R}{>{\hsize=0.9\hsize\raggedleft\arraybackslash}X}

\algnewcommand{\algorithmicsubalgorithm}{\textbf{function}}
\algdef{SE}[SUBALG]{SubAlgorithm}{EndSubAlgorithm}[1]{\algorithmicsubalgorithm\ \textsc{#1}}{\algorithmicend\ \algorithmicsubalgorithm}
\algtext*{EndSubAlgorithm}
\algnewcommand{\algorithmicinput}{\textbf{Input:}}
\algnewcommand{\algorithmicoutput}{\textbf{Output:}}
\algnewcommand\Input{\item[\algorithmicinput]}%
\algnewcommand\Output{\item[\algorithmicoutput]}%
\onlineid{1417}

\vgtccategory{Research}
\vgtcpapertype{evaluation}

\title{Causal Support: Modeling Causal Inferences with Visualizations}

\author{Alex Kale, Yifan Wu, and Jessica Hullman}
\authorfooter{
\item
Alex Kale is with the University of Washington. E-mail: kalea@uw.edu.
\item
Yifan Wu is with the University of California at Berkeley. E-mail: yifanwu@berkeley.edu.
\item
Jessica Hullman is with Northwestern University. E-mail: jhullman@northwestern.edu.
}

\shortauthortitle{Kale \MakeLowercase{\textit{et al.}}: Causal Support: Modeling Causal Inferences with Visualizations}
\abstract{
Analysts often make visual causal inferences about possible data-generating models. However, visual analytics (VA) software tends to leave these models implicit in the mind of the analyst, which casts doubt on the statistical validity of informal visual ``insights''.
We formally evaluate the quality of causal inferences from visualizations by adopting \textit{causal support}---a Bayesian cognition model that learns the probability of alternative causal explanations given some data---as a normative benchmark for causal inferences.
We contribute two experiments assessing how well crowdworkers can detect (1) a treatment effect and (2) a confounding relationship.
We find that chart users’ causal inferences tend to be insensitive to sample size such that they deviate from our normative benchmark.  
While interactively cross-filtering data in visualizations can improve sensitivity, on average users do not perform reliably better with common visualizations than they do with textual contingency tables.
These experiments demonstrate the utility of causal support as an evaluation framework for inferences in VA and point to opportunities to make analysts' mental models more explicit in VA software.
} % end of abstract


\CCScatlist{ % not used in journal version
 \CCScat{H.m}{Information Systems}%
{Miscellaneous}
}

\teaser{
  \centering
  \includegraphics[width=\linewidth]{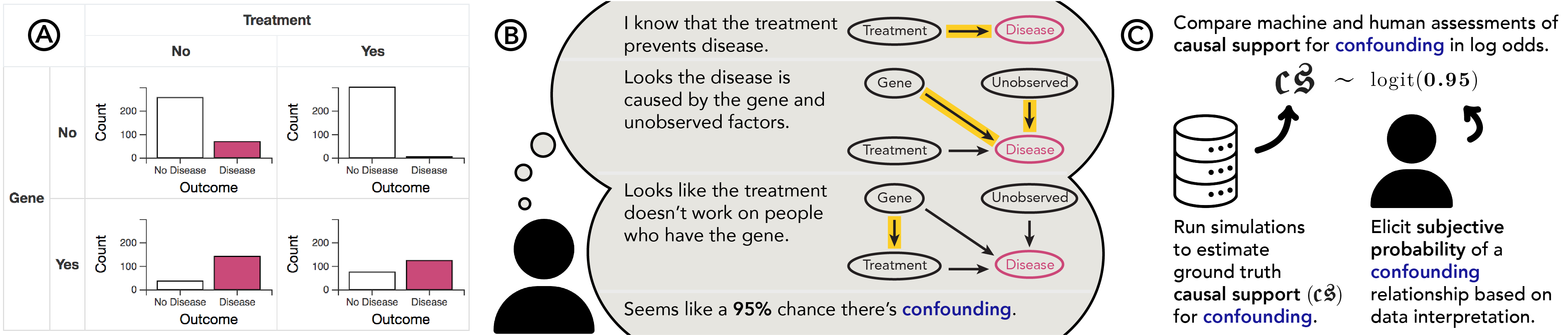}
  \setlength{\abovecaptionskip}{-10pt}
  \setlength{\belowcaptionskip}{-5pt}
  \caption{Modeling causal inferences with visualizations: \raisebox{.5pt}{\textcircled{\raisebox{-.6pt} {A}}} Users view and may interact with data visualizations; \raisebox{.5pt}{\textcircled{\raisebox{-.8pt} {B}}} Ideally, users reason through a series of comparisons that allow them to allocate subjective probabilities to possible data generating processes; and \raisebox{.5pt}{\textcircled{\raisebox{-1pt} {C}}} We elicit users' subjective probabilities as a Dirichlet distribution across possible causal explanations and compare these causal inferences to a computed benchmark of causal support, which we derive from Bayesian inference across possible causal models.
  } 
  \label{fig:teaser}
}

\vgtcinsertpkg

\begin{document}


\firstsection{Introduction}

\maketitle

%% \section{Introduction} %for journal use above \firstsection{..} instead

% \alex{Hook: unpack the relationship between visualization and causal inference. What is the utility of an external representation?}
Data analysts engaged in sensemaking~\cite{Card1999,Pirolli2005,Russell1993,Grolemund2014} infer the compatibility of different causal explanations with their data.
During exploratory analysis or provisional statistical modeling, analysts implicitly or explicitly compare their data to patterns they expect as logical consequences of hypothesized data generating processes~\cite{Andrienko2018,gelman2003,gelman2004exploratory,Hullman2020,hullman2020theories}.
Visualizations play a critical role in causal inference both because externalization reduces cognitive load~\cite{Cox1999} and because human capabilities for inference rely heavily on sensory expectations (e.g., mental imagery) and comparisons between expectations and experiences~\cite{Breedlove2020,Kale2019adaptation}.

% \alex{Talk about the ways in which causal inference can be error prone. Motivate why it is important to model causal inferences with visualizations. We need an evaluative approach to better integrate generative models into visual data analysis software.}
Data analysts and software designers need to anticipate how human capabilities for causal inference may be error prone.
For instance, perceptual biases such as underestimation of sample size (e.g.,~\cite{Kim2019}) contribute to errors in causal inferences insofar as perceived associations seem to be the basis for causal inferences~\cite{Anderson1995,Xiong2019}.
Analysts also err in their causal interpretations of data when the mapping between a potential causal explanation and an expected pattern in the data is unclear~\cite{Batanero1996,Yen2019}.
For example, imagine an analyst trying to detect confounding in experiment results on the effectiveness of a treatment at preventing disease (Fig.~\ref{fig:teaser} \raisebox{.5pt}{\textcircled{\raisebox{-.6pt} {A}}}). To detect that `gene' is a confounding factor, the analyst must see effects of gene on both treatment effectiveness (i.e., a difference between the top and bottom cells in the right column of table \raisebox{.5pt}{\textcircled{\raisebox{-.6pt} {A}}}) and overall rate of disease (i.e., a difference between the top and bottom rows of \raisebox{.5pt}{\textcircled{\raisebox{-.6pt} {A}}}). 
Attributing these signals in the data to confounding requires the analyst to know what they are looking for,
% rather than assuming they will passively detect the appropriate visual patterns as insights.
rather than passively detecting the appropriate patterns.

% How well do visualizations common in visual analytics software support causal inferences about possible data generating processes?
% Does the ability to interactively filter or aggregate data help analysts discover causal relationships?
% To answer these questions, we need to compare analysts' causal inferences to a benchmark that is roughly `normative', that captures important aspects of good causal inference.

% \alex{Talk about what we did and what we found}
To assess how well visual analytics (VA) tools support such causal inferences, we need to compare analysts' inferences to a benchmark that is roughly `normative', that captures important aspects of good causal inference.
% In this study, w
We adopt causal support, a model from mathematical psychology~\cite{Griffiths2005}, as a benchmark for evaluating causal inferences from visualizations. 
\textit{Causal support} models one's belief in a set of
% alternative 
possible data generating processes as a Bayesian update.
% We posit c
Causal support is a good normative model for three reasons. 
(1) Causal support has numerous proprieties of valid statistical inference in light of the analyst's prior knowledge. It captures the fact that belief in evidence 
% from the data 
should be stronger as sample size increases. It accounts for unknown unknowns about the space of possible models, such that causal support assigns no posterior probability to models that the analyst does not explicitly consider. 
(2) Prior work~\cite{Griffiths2005} shows through a system of experiments that causal support accounts for otherwise hard-to-explain patterns in human causal inferences,
e.g.,
% For example, causal support correctly predicts 
that subjective belief in a causal relationship varies as a function of the potential to detect that relationship in a given data set.
(3) Since causal support is \textit{extensible to any generative model} 
% (i.e., models that produce predictions and can assign likelihood to data), this evaluation approach can be applied in a wide range of visual analytics applications to evaluate the quality of causal inferences. 
(i.e., models that can assign likelihood to data), it can be applied in a wide range of 
VA
% visual analytics 
applications to evaluate 
% the quality of 
causal inferences.

% \alex{Unpack contributions}
% We contribute two experiments using causal inference problems involving count and proportion data to (1) study the utility of causal support for gaining insight into inferences from visualizations and (2) evaluate how well visualizations common in visual analytics (VA) software support causal inferences.
We contribute two experiments using causal inference problems involving count and proportion data to (1) study the utility of causal support for gaining insight into inferences from visualizations and (2) evaluate how well visualizations common in visual analytics (VA) software support causal inferences about possible data generating processes.
We compare three common visual encodings for count and proportion data---bar charts, icon arrays, and text tables as a baseline---and we investigate how the ability to interactively aggregate or cross-filter data in bar charts impacts causal inferences. 
In Experiment 1, we ask participants to differentiate whether a treatment is effective by allocating probability across two different data generating processes.
We find that chart users' causal inferences are far from normative with all visualizations we tested, and interacting with visualizations can improve sensitivity to signal in predictable ways. 
Ultimately, however, no visualization reliably outperforms text tables. 
We also see that chart users are more sensitive to evidence against a treatment effect than evidence in favor of one, suggesting an unequal weighting of falsifying versus verifying evidence.
% In Experiment 2, we replicate the main findings Experiment 1 but with a task where we ask participants to detect confounding in data sets by allocating probability across four alternative data-generating processes. 
In Experiment 2, we replicate the main findings Experiment 1 but with a confounding detection task where participants allocate probability across four alternative data generating processes. 
% The models in this second experiment demonstrate how casual support can be extended to study complex causal inferences across many models. 
Experiment 2 demonstrates how casual support can be extended to study causal inferences about more complex data generating processes.
\section{Related Work}
\subsection{Visualization for casual inference}
Much of the psychology and statistics literature on visual aids for causal reasoning focuses on contingency tables (e.g.,~\cite{Anderson1995,Batanero1996,Cheng1997,Greenland1999,Griffiths2005,Sobel1995}).
Contingency tables support causal inferences by using layout to encode conditional probabilities, the same way trellis plots afford grouping by factors during visual data analysis~\cite{Tukey1977,Becker1996}.
Whether or not a factor seems to be \textit{collapsible}---whether or not patterns the data seem to change depending on whether the data are grouped by that factor---can be a visual signal for reasoning about causal relationships such as confounding~\cite{Greenland1999}.
However, empirical research on interpretation strategies for contingency tables~\cite{Batanero1996} suggests that analysts often misinterpret signals like collapsability because they don't ascertain the mapping between 
% visual signals like collapsability 
these visual signals
and hypothesized causal relationships.
Tools like Tableau enable users to explore collapsability by interactively grouping data.
% Tools like Tableau support exploration of collapsability by enabling analysts to interactively control how data are aggregated.
We investigate whether the ability to interactively control data aggregation improves the quality of causal inferences.
% , showing how strategic errors in causal reasoning impact the use of interactive visualizations.

Research on visual analytics (VA) employs a broader range of representations to support causal reasoning, including parallel coordinates~\cite{Wang2016,Wang2018}, bar charts~\cite{Yen2019}, ``diff bar charts'' showing counterfactual outcomes under different conditions as layered bars~\cite{Xie2020preprint}, and novel techniques using animation to show event sequences (e.g.,~\cite{Elmqvist2003,Jin2020,Kadaba2007}).

Some of these tools also incorporate directed acyclic graphs (DAGs) as interfaces to models and visualizations (e.g.,~\cite{Wang2016,Wang2018,Xie2020preprint,Yen2019}).
DAGs are 
% a visual tool
devices for causal reasoning which have garnered attention in recent years~\cite{Pearl2018}. 
DAGs encode hypothesized relationships among variables (e.g.,~Fig.\ref{fig:teaser}\raisebox{.5pt}{\textcircled{\raisebox{-.8pt} {B}}}), making causal relationships and the assumptions they entail explicit and in some cases testable~\cite{Bareinboim2016,Pearl2009,Pearl2014,Pearl2015}.
We use DAGs to present differences between alternative causal explanations for data sets that we ask participants to judge (Figs.~\ref{fig:dags-e1-explainer}~\&~\ref{fig:dags-e2-explainer}).
% In this study, we use DAGs to present alternative causal explanations for data sets that we ask participants to judge as a way of making these alternatives and the differences between them explicit (Figs.~\ref{fig:dags-e1-explainer}~\&~\ref{fig:dags-e2-explainer}).

VA systems frequently use interaction techniques such as cross-filtering linked views of data (e.g.,~\cite{Wang2016,Wang2018}) and click- or drag-and-drop-based chart construction (e.g.,~\cite{stolte2002polaris,Xie2020preprint,Yen2019}).
The most similar prior research to our study\footnote{\mbox{Also see \url{https://logical-interactions.github.io/causal2020/}}
% unpublished formative work from the second author: 
% \\ \texttt{URL redacted for anonymous review}
} tests whether constructing charts by clicking on variables versus dragging variables onto a DAG makes a difference in analysts' ability to differentiate between different kinds of causal relationships~\cite{Yen2019}, specifically identifying mediating variables.
Although they do not find an effect of interaction method on causal inferences, the authors provide a detailed strategy analysis extending evidence from psychological studies~\cite{Batanero1996} that analysts struggle to reason about the exact set of visual signals they should look for to verify or falsify a causal relationship.
We extend this line of work by studying whether the ability to (un)facet charts or cross-filter coordinated multiple views impacts the quality of untrained analysts' causal inferences.

Prior work in visualization~\cite{Micallef2012,Ottley2016} and risk communication~\cite{Galesic2009,Spiegelhalter1999} suggests that icon arrays can improve Bayesian inferences,
perhaps
% This may be 
in part because of cognitive benefits of framing probabilities as frequencies of events~\cite{Gigerenzer1995,Hoffrage1998,Hullman2015,Kay2016}.
% However, the performance improvements associated with these visualizations seem to depend on the use of text in the display~\cite{Micallef2012}, the chart user's spatial ability~\cite{Ottley2016}, and the strategies users rely on~\cite{Kale2021}.
We compare icon arrays to text tables and bar charts since these visualizations 
span
% represent very different parts of 
the design space for showing count data and are also easy to create in 
VA
% visual analytics 
software like Tableau.

\subsection{Modeling causal reasoning}
In the present study, we draw on and extend a model of causal reasoning called causal support, first proposed by Griffiths and Tenenbaum~\cite{Griffiths2005}.
\textit{Causal support} formulates causal inferences as a Bayesian update on the log odds of a finite set of causal explanations given some observed data. 
Mathematically, causal support has similar properties to a Chi-squared test (i.e., Are the data in each cell of a contingency table likely generated by the same process?), which prior work analogizes to the kind of comparisons between data and model predictions that analysts visualize in ``model checks''~\cite{gelman2003,gelman2004exploratory,hullman2020theories} such as QQ-plots.
However, unlike a Chi-squared test, causal support relies on Monte Carlo simulations to assign likelihoods under alternative causal explanations, making causal support extensible to any finite set of generative causal models.
For instance, Pacer and Griffiths extend causal support to handle continuous data~\cite{Pacer2011} and event streams~\cite{Pacer2015}.
Similarly, in Experiment 2, we present 
an
% a novel 
extension of causal support to evaluate inferences about more than two possible data generating models.

Previous cognitive models of causal inference explored in psychology share more in common with parameter estimation than statistical inference per se, a subtle but important distinction.
One such model \textit{delta p} posits that that people judge differences in conditional proportions of observed events when making causal inferences about count data~\cite{Anderson1995}.
Another such model \textit{causal power} posits that people judge the magnitude of effect size when making causal inferences~\cite{Cheng1997}. 
% Both of these predecessors and alternatives to causal support make the assumption that causal inferences are fundamentally a perceptual judgment.
% The difference between them is that causal power rescales delta p to apportion the posited signal for causal inferences based on the potential to detect any signal whatsoever within the observed data.
Both of these predecessors to causal support make the assumption that causal inferences are fundamentally a perceptual judgment, however, causal power rescales delta p based on the potential to detect any signal whatsoever within the observed data.
In contrast, causal support assumes that the signal for causal inferences depends on the possible data generating models that the analyst has in mind and represents these alternative models explicitly.
This makes causal support more flexible, with higher predictive validity for human judgments than delta p, causal power, and even Chi-squared~\cite{Griffiths2005}. 
Causal support 
% models structure learning in a way that 
reflects analysts' natural tendency to dichotomize, for better or worse, reasoning about whether or not causal relationships exist rather than how strong they are.
\section{Experiment 1}
\textit{How well do different visualization designs that are common in visual analytics (VA) software support causal inferences about possible data generating processes?} 
We evaluate visualizations of count data including text contingency tables, icon arrays, grouped bar charts, bars that users can interactively aggregate, and linked bars that users can interactively cross-filter.
In Experiment 1, we investigate chart users' ability to infer whether a treatment prevents a disease.
By asking chart users about a treatment effect in count data, we build on the task and structural equation models used by Griffiths and Tenenbaum~\cite{Griffiths2005} to propose and validate causal support.
Count data are also ideal for evaluating bar charts.
The design requirements for supporting our causal inference task with count data are that visualizations should express both the \textit{proportion} of people with disease and \textit{sample size}.
% Based on these requirements, we ruled out testing visualizations such as pie charts and heatmaps, which are common ways of encoding proportions but which don't encode sample size.
Based on these requirements, we ruled out testing pie charts and heatmaps, common ways of encoding proportions that do not encode sample size.

\subsection{Method}
We set out to study how well different visualizations support causal reasoning by using \textit{causal support} as a benchmark for causal inferences.

\subsubsection{Task scenario \& response elicitation}
Participants played the role of an analyst hired by a company to interpret samples of data on the effectiveness of experimental treatments at preventing various diseases.
We showed participants visualizations of the number of people in each sample who did or did not receive \textit{treatment}, get a \textit{disease}, and have a \textit{gene} known to cause the disease.

\begin{figure}[t]
    \centering
    \includegraphics[width=\columnwidth]{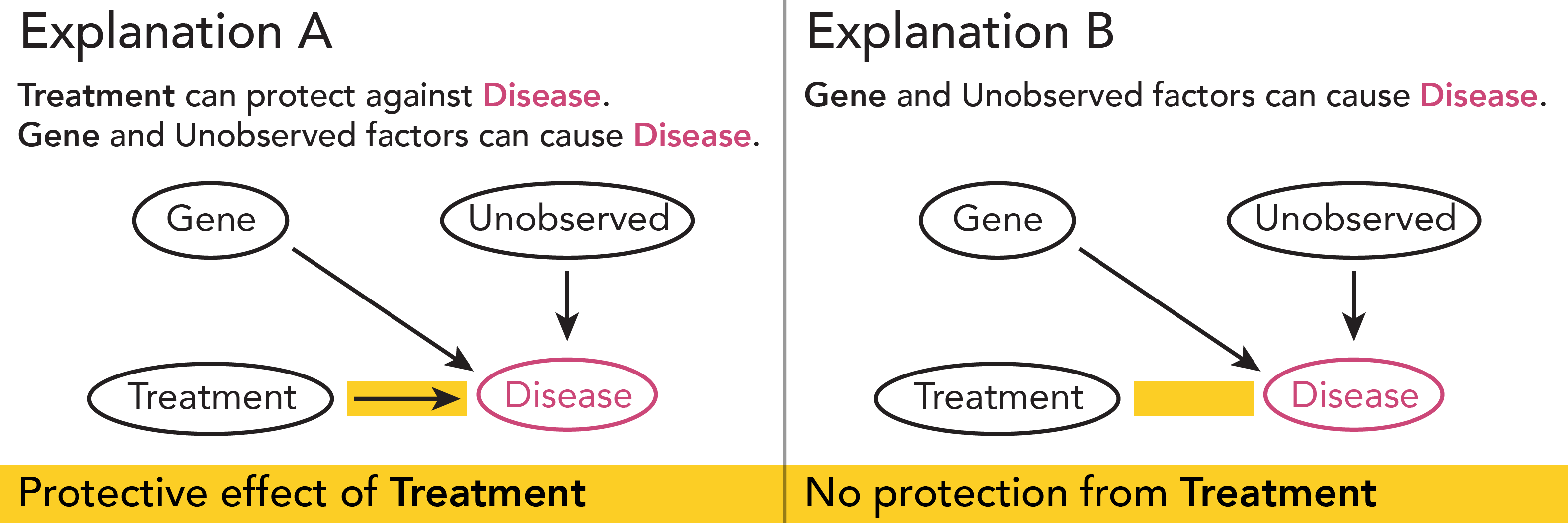}
    \setlength{\abovecaptionskip}{-10pt}
    \setlength{\belowcaptionskip}{0pt}
    \caption{DAGs representing possible causal explanations participants were asked to consider in Experiment 1.
    }
    \label{fig:dags-e1-explainer}
\end{figure}

We asked participants to judge the underlying causal relationships in the data, rating their degree of belief that treatment protects against disease by allocating probability across the two DAGs in Figure~\ref{fig:dags-e1-explainer}.
We chose to study causal inferences about treatments, genes, and diseases in order to create a scenario where users would find the possible causal explanations feasible, coherent, and memorable.

\noindent
\textbf{Question \& elicitation.}
We asked participants the following question: 
\begin{quote}
    \vspace{-7pt}
    \textit{\textbf{How much do you believe in each of the causal explanations described below?} Imagine you have 100 votes to allocate across the two possible explanations. Split your 100 votes between explanations based on your degree of belief. For example, if you think one explanation is twice as likely as the other, you might give 67 votes (roughly two thirds) to that explanation and 33 votes (roughly one third) to the other. Assume no other explanations are possible.}
\end{quote}
\vspace{-7pt}
\noindent
Participants responded with two complementary probabilities. We used form validation to make sure their responses were both numbers between 0 and 100 that summed to 100.
Following prior work on eliciting Dirichlet distributions~\cite{Chalone1987,OHagan2006} (i.e., probabilities allocated across alternatives), when participants gave their first response, we imputed what the second response would need to be in order for their responses to sum to 100. This imputed value and a corresponding prompt, \textit{``Adjust your responses until both numbers reflect your beliefs.''}, were both highlighted with the same color to indicate this imputation.
We elicited probabilities as ``votes out of 100'' because \textit{frequency framing} tends to reduce bias in probability estimates~\cite{Gigerenzer1995,Hoffrage1998,OHagan2006}.
Participants received no feedback on their responses.
% We combined these subjective probabilities  ($\mathit{response_A}$ and $\mathit{response_B}$) into an estimate of perceived causal support, which we compared to our normative benchmark. 
We transformed these responses into perceived causal support, which we compared to our benchmark.

\begin{figure}[b]
    \centering
    \includegraphics[width=\columnwidth]{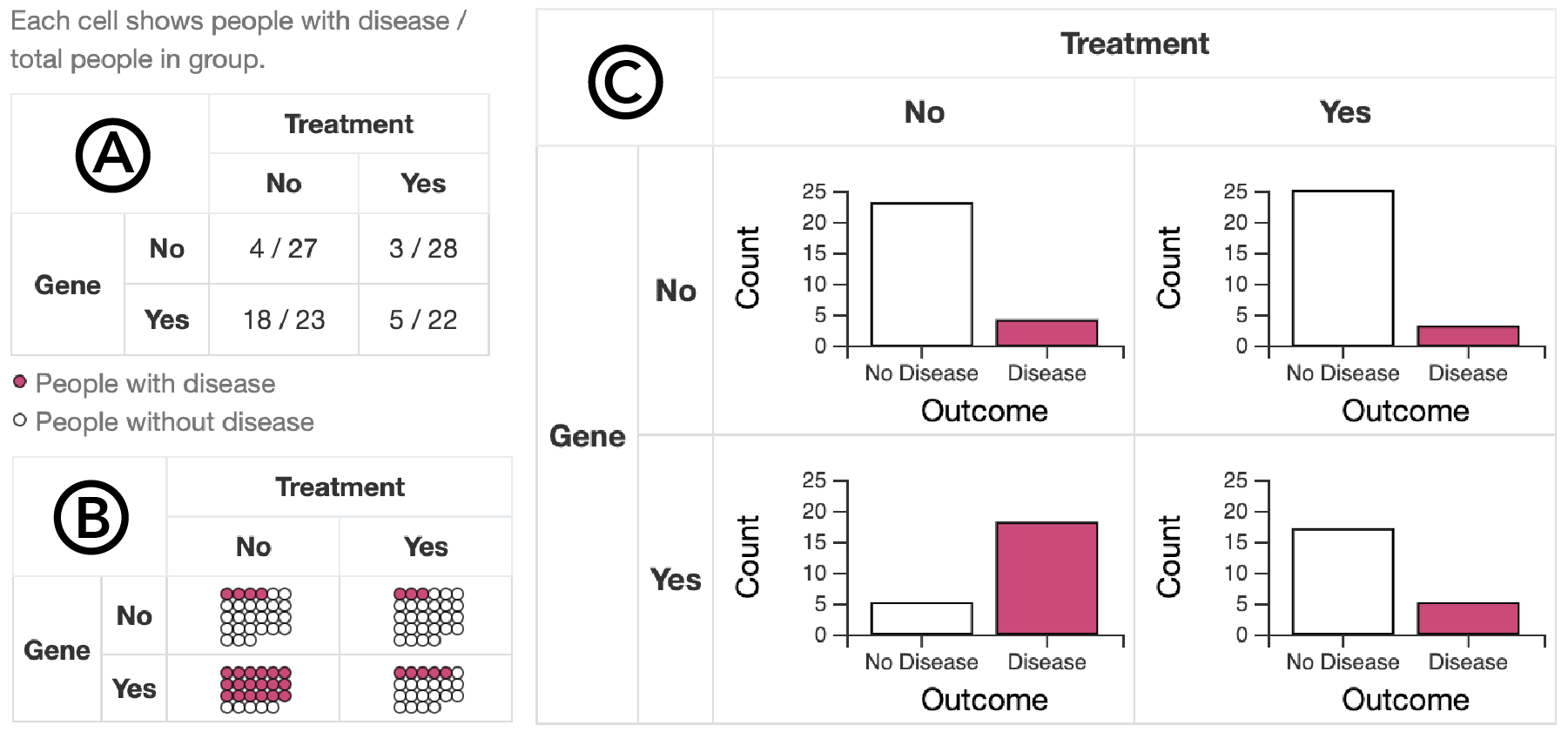}
    \setlength{\abovecaptionskip}{-10pt}
    \setlength{\belowcaptionskip}{0pt}
    \caption{
        Non-interactive visualizations evaluated in our study: \raisebox{.5pt}{\textcircled{\raisebox{-.6pt} {A}}} text contingency tables; \raisebox{.5pt}{\textcircled{\raisebox{-.8pt} {B}}} faceted icon arrays; and \raisebox{.5pt}{\textcircled{\raisebox{-1pt} {C}}} faceted bar charts.
    }
    \label{fig:non-interactive-vis-conds}
\end{figure}

\noindent
\textbf{Perceived causal support.}
The dependent variable in our study was a measure of the perceived log odds of a \textit{target explanation} over other possible causal explanations.
Specifically, in Experiment 1 we targeted explanation A, which posited a treatment effect, requiring us to transform participants' responses into a log response ratio ($\mathit{lrr_A}$),
\begingroup
\setlength\abovedisplayskip{3pt}
\setlength\belowdisplayskip{3pt}
    $$\mathit{lrr_A} = \log \bigg( \frac{\mathit{response_A}}{\mathit{response_B}} \bigg)$$
\endgroup
where $\mathit{response_A}$ and $\mathit{response_B}$ were the probabilities participants allocated to causal explanations A and B, respectively, on each trial.
We used a log odds scale in order to make participants' perceived causal support comparable to our normative benchmark of causal support.

\noindent
\textbf{Payment.}
% Participants received a \textit{guaranteed reward of \$2 for participating} plus a bonus of \$0.25 for every trial where their estimate of the probability of causal explanation A was within 5 votes of the ground truth derived from our benchmark model of causal support.
Participants received a \textit{guaranteed reward of \$2} plus a bonus of \$0.25 for every trial where their estimate of the probability of causal explanation A\footnote{Bonuses in Experiment 2 were based on the probability of explanation D.} was within 5 percentage points of the ground truth.

\noindent
\textbf{Apparatus.}
We collected data using a Flask application deployed on Heroku with a Firebase database and visualizations created with D3.\footnote{
% \url{http://causal-support.herokuapp.com/0_landing?workerId=dev&assignmentId=test&batch=999&cond=filtbars}
E.g., see Experiment 2 interface at \url{https://bit.ly/3rDcxfn}
}

% \begin{wrapfigure}{r}{0.4\columnwidth}
%     \vspace*{-4mm}
%     \centering
%     \includegraphics[width=0.4\columnwidth]{figures/vis-conditions_text.png}
%     \setlength{\abovecaptionskip}{-7pt}
%     \setlength{\belowcaptionskip}{-7pt}
%     \caption{Text tables.}
%     \label{fig:text}
    
%     \vspace*{4mm}
%     \centering
%     \includegraphics[width=0.4\columnwidth]{figures/vis-conditions_icons.png}
%     \setlength{\abovecaptionskip}{-7pt}
%     \setlength{\belowcaptionskip}{-7pt}
%     \caption{Icon arrays.}
%     \label{fig:icons}
    
%     % \vspace*{3mm}
%     % \centering
%     % \includegraphics[width=0.4\columnwidth]{figures/vis-conditions_bars.png}
%     % \setlength{\abovecaptionskip}{-5pt}
%     % \setlength{\belowcaptionskip}{-7pt}
%     % \caption{Bar charts.}
%     % \label{fig:bars}
    
%     % \vspace*{3mm}
%     % \centering
%     % \includegraphics[width=0.4\columnwidth]{figures/vis-conditions_aggbars.png}
%     % \setlength{\abovecaptionskip}{-12pt}
%     % \setlength{\belowcaptionskip}{-12pt}
%     % \caption{Aggregating bars.}
%     % \label{fig:aggbars}
    
%     % \vspace*{3mm}
%     % \centering
%     % \includegraphics[width=0.4\columnwidth]{figures/vis-conditions_filtbars.png}
%     % \setlength{\abovecaptionskip}{-5pt}
%     % \setlength{\belowcaptionskip}{-7pt}
%     % \caption{Cross-filter bars.}
%     % \label{fig:filtbars}
% \end{wrapfigure}

\subsubsection{Visualization conditions}
Our visualizations show the number of people with and without disease in each cell of a 2x2 contingency table faceted by treatment and gene, with the exception of cross-filter bars which use a different layout.
We aimed to test visualizations of count data similar to what an analyst could produce using visual analytics software like Tableau.

\noindent
\textbf{Text contingency tables}
% The text condition 
showed the number of people with disease as a fraction of the total number of people in each cell of a faceted table (Fig.~\ref{fig:non-interactive-vis-conds}~\raisebox{.5pt}{\textcircled{\raisebox{-.6pt} {A}}}).
Text tables, which have been studied in prior 
% psychology 
research on causal support, served as a baseline comparison for other visualizations.

\noindent
\textbf{Icon arrays}
% Icon arrays 
showed counts of people with and without disease as filled and open circles, respectively (Fig.~\ref{fig:non-interactive-vis-conds}~\raisebox{.5pt}{\textcircled{\raisebox{-.8pt} {B}}}). 
We set the number of dot columns to minimize the aspect ratio on each trial, similar to how analysts might create roughly square icon arrays in Tableau.
Icon arrays express both proportion and sample size as natural frequencies, 
% an information format 
which prior work finds beneficial for statistical reasoning (e.g., ~\cite{Galesic2009, Gigerenzer1995,Hoffrage1998}).

\noindent
\textbf{Bar charts}
% Grouped bar charts 
showed counts of people with and without disease using a length/position encoding on a common scale (Fig.~\ref{fig:non-interactive-vis-conds}~\raisebox{.5pt}{\textcircled{\raisebox{-1pt} {C}}}).
On each trial, we set the $y$-axis scale to the maximum count of the data in view, allowing scales to change from trial to trial as they do when users load a new data set in Tableau. %We made the same design choice with the axis scales in every variation of bar charts (see below). 
% Grouped
Bar charts are ubiquitous for 
% showing 
count data.

% \begin{wrapfigure}{r}{0.4\columnwidth}
%     \vspace*{-7mm}
%     \centering
%     \includegraphics[width=0.4\columnwidth]{figures/vis-conditions_aggbars.png}
%     \setlength{\abovecaptionskip}{-12pt}
%     \setlength{\belowcaptionskip}{-12pt}
%     \caption{Aggregating bars.}
%     \label{fig:aggbars}
% \end{wrapfigure}

\begin{figure}[t]
    % \vspace*{-3mm}
    \centering
    \includegraphics[width=\columnwidth]{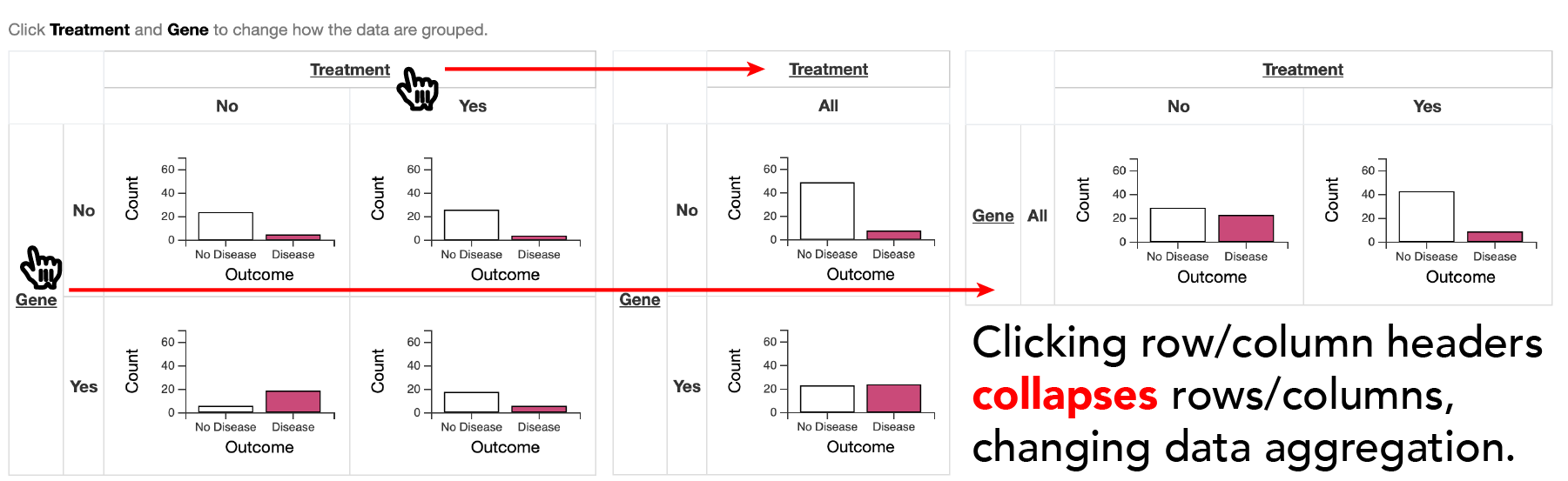}
    \setlength{\abovecaptionskip}{-10pt}
    \setlength{\belowcaptionskip}{0pt}
    \caption{Aggregating bars mimic shelf construction and faceting.
    }
    \label{fig:aggbars}
\end{figure}

\begin{figure}[b]
    % \vspace*{-3mm}
    \centering
    \includegraphics[width=\columnwidth]{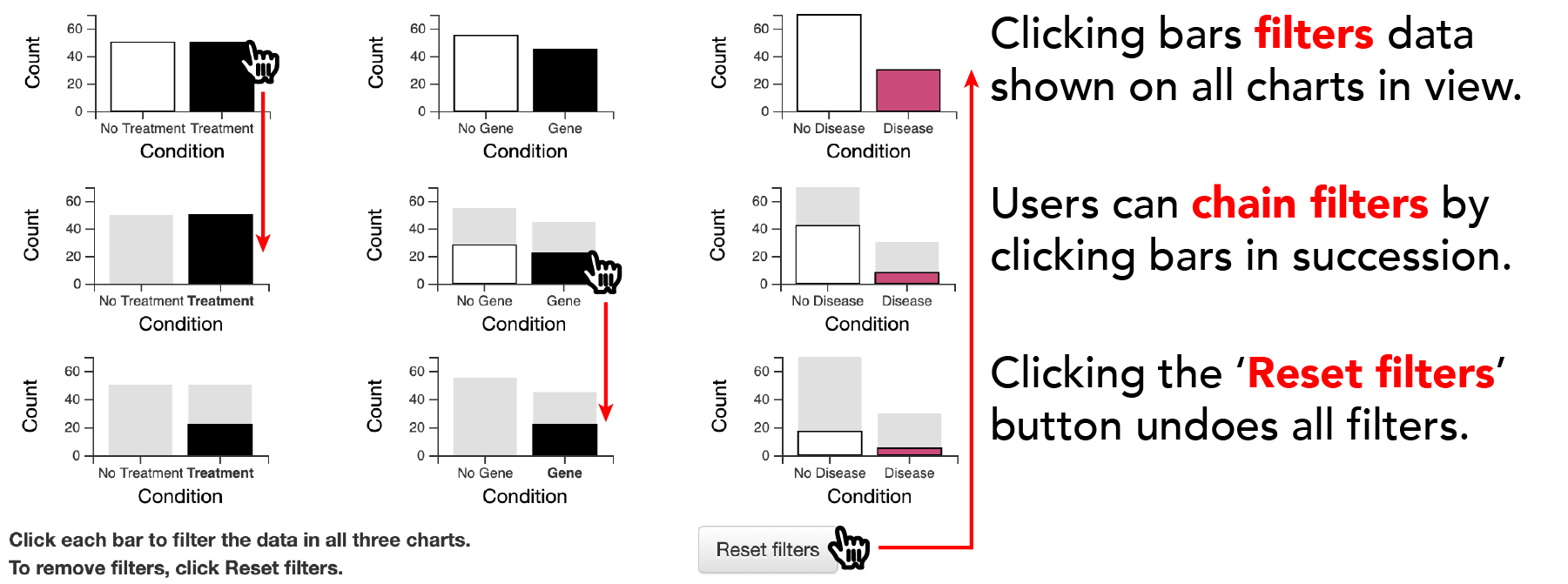}
    \setlength{\abovecaptionskip}{-10pt}
    \setlength{\belowcaptionskip}{0pt}
    \caption{Cross-filtering bars mimic coordinated multiple views.
    }
    \label{fig:filtbars}
\end{figure}

\noindent
\textbf{Aggregating bars}
% Aggregating bars 
(aggbars) were similar to bar charts, however, users could interactively
% control data aggregation by clicking on the table headers to toggle faceting by treatment or gene (Fig.~\ref{fig:aggbars}).
toggle faceting by treatment or gene by clicking on the table headers (Fig.~\ref{fig:aggbars}).
% To accommodate interactive aggregation,
On each trial, we set the $y$-axis scale to the maximum count of the fully aggregated data.
We designed aggbars to roughly mirror Tableau's shelf interactions, where users control faceting by direct manipulation of table headers.
Interactive faceting may facilitate causal inferences by enabling users to explore whether ``collapsing''~\cite{Greenland1999} over a factor changes patterns in the data.

\noindent
\textbf{Cross-filtering bars}
were three bar charts showing the number of people with and without treatment, the gene, and disease, respectively (Fig.~\ref{fig:filtbars}).
We linked these bar charts such that clicking on one bar cross-filtered the rest of the data in view.
When users applied a filter 
% (e.g., to only show data for people who received treatment), 
(e.g., only show people who received treatment), 
the corresponding axis label became bold and 
% to signal which filters the user has applied.Additionally, 
gray bars persisted in the background,
% behind the filtered bars 
% so that chart users do not need to rely on working memory to compare filtered and unfiltered views. 
so chart users could compare filtered and unfiltered views without relying on working memory.
Users ``reset filters'' by clicking a button below the charts.
% Unlike our other visualizations, cross-filtering bars did not use a two-way table layout. 
% We implemented three linked bar charts to 
Filterbars emulated coordinated multiple views such as 
% analysts might create in 
a Tableau dashboard.
Interactive cross-filtering might assist in causal inferences because conditioning the data in view based on a specific event is analogous to Pearl's \textit{do operator} (e.g.,~$\mathrm{do}(\mathit{treatment}=\mathrm{yes})$), a notation for 
% expressing and reasoning with counterfactual interventions
reasoning about counterfactuals in a causal networks~\cite{Pearl2018}.

\subsubsection{Experimental design}
We manipulated both the visualizations participants used for the task and the data sets that we showed.
We randomly assigned each participant to use one of five visualizations (see Section 3.1.2), making comparisions of visualizations between-subjects.
We showed each participant a total of 18 trials, which included 16 data conditions (see Section 3.1.4) presented within-subjects and two attention checks which we used for exclusion criteria (see Section 3.1.6).
We randomized trial order for each participant, inserting attention checks on trials 7 and 13.

\subsubsection{Stimulus generation}
We evaluated causal inferences on realistic data sets, which spanned a range of ground truth causal support.
Generating data sets required (1) manipulating data attributes which signaled causal support to participants, and (2) labeling each data set with ground truth causal support. 
% via a Monte Carlo simulation.

Our goal was to generate 16 data conditions (i.e., trials in our experiment) that varied delta p and sample size, two data attributes which in turn manipulate ground truth causal support.
\textit{Delta p} described the difference in the proportion of people with disease in each data set depending on whether they received treatment.
Positive values of delta p indicated evidence that the treatment protected against disease; negative values indicated evidence against treatment effectiveness.
\textit{Sample size} was the number of people in each data set we showed participants.

\noindent
\textbf{Data conditions.}
To generate our 16 data conditions, we simulated data from structural models with one parameter per DAG arrow in Figure~\ref{fig:dags-e1-explainer}.
We manipulated both the \textit{probability that treatment prevents disease} (4 levels: $\{0, 0.1, 0.2, 0.4\}$) and \textit{sample size} (4 levels: $\{100, 500, 1000, 1500\}$).
We controlled the probability of disease due to the gene ($0.5$), probability of disease due to unobserved causes ($0.2$), base rate of the gene ($0.4$), and the proportion of each sample with treatment ($0.5$).
We selected these parameters iteratively by sampling data sets and labeling ground truth until half of the trials had greater than a 50\% chance of being generated by causal explanation A.
% \jessica{I think the detail about the data conditions needs to come earlier, before we describe data generation process. Can just be a single sentence, like Our goal was to generate 16 data conditions (which become trials in our experiment) that varied two factors: .... This will give some needed context on our goals in the descdription of the dataset generation. Actually should probably just describe data generation with modeling causal support. Otherwise its very hard to figure out how they go together}

For each of the resulting 16 data conditions, we simulated many data sets using a binomial random number generator to approximate realistic sampling error.
By simulating sampling error, we prevented the count data from appearing contrived.
% Because of this sampling error, ground truth causal support was not a deterministic function of the factors we manipulated to generate the data; \jessica{for small sample sizes in particular, the computed causal support could be fairly distinct from that implied if we had not simulated sampling error [is this true? trying to give some intuition]}.
% To guarantee that each participant saw trials spanning a consistent range of causal support despite this sampling error, we counterbalanced the presentation of this sampling error across participants within each visualization condition. \jessica{this is probably the hardest part of this to understand. im not quite sure what it means to counterbalance the sampling error}
% We selected a subset of data sets for each data condition representing 16 quantiles of causal support, and we counterbalanced which quantile was shown for each data condition using a balanced latin square.
This sampling error resulted in a distribution of ground truth causal support under each data condition, with more variability  in the ground truth at smaller sample sizes. 
To guarantee that each participant saw trials spanning a consistent range of causal support, we selected 16 data sets representing 16 quantiles of the ground truth distribution per data condition, and we counterbalanced the quantile shown for each data condition across participants within each visualization condition using a balanced latin square.
For our attention check trials, we selected the two simulated data sets that had the minimum and maximum ground truth causal support.

\begin{algorithm}[t]
    \caption{
        Monte Carlo simulation to calculate causal support in Experiment 1. Algorithm for Experiment 2 is similar.
    }
    \label{alg:monte_carlo}
\begin{algorithmic}[1]
    \Input (8, 1) vector of contingency table counts (no disease vs disease, no gene vs gene, no treatment vs treatment) $C$, Monte Carlo iterations $m$, set of parameters to fix at zero $\theta_0$ (i.e., parameters representing DAG arrows to omit from the data generating process)
    \Output \textsc{{monte\_carlo}} returns log likelihood of the given data generating process $\mathit{log\_lik}$; Main returns causal support for the target explanation (Fig.~\ref{fig:dags-e1-explainer}, Explanation A) $\mathfrak{cs}_A$ 
    \item \textcolor{red}{\texttt{\textbf{\# Monte Carlo simulation to calculate likelihood}}}
    \SubAlgorithm{monte\_carlo}{($C$, $m$, $\theta_0$):} \\
        \STATE \hspace{\algorithmicindent} Parameters corresponding to each DAG arrow in Fig.~\ref{fig:dags-e1-explainer}: \\
        \hspace{\algorithmicindent} $\theta_\mathcal{P} = \{$ \textcolor{red}{\texttt{\ \ \ \ \ \# initialize parameters}} \\
        \hspace{\algorithmicindent} \hspace{\algorithmicindent}  $\Pr(\mathcal{D})$, \textcolor{red}{\texttt{\ \ \ \# p disease due to unknown causes}} \\
        \hspace{\algorithmicindent} \hspace{\algorithmicindent} $\Pr(\mathcal{D}|G)$, \textcolor{red}{\texttt{\ \# p disease due to gene}} \\
        \hspace{\algorithmicindent} \hspace{\algorithmicindent} $\Pr(\neg \mathcal{D}|T)$ \textcolor{red}{\texttt{\# p no disease due to treatment}} \\
        \hspace{\algorithmicindent} $\}$
        \For{parameter $\theta \in  \theta_\mathcal{P}$} \textcolor{red}{\texttt{\ \# assign parameters}}
            \If{$\theta \in \theta_0$} Fix parameter at zero: $\theta = \mathrm{Zeros(}m\mathrm{)}$
            \Else{} Uniformly sample probabilities: $\theta = \mathrm{Random(}0,1,m\mathrm{)}$
            \EndIf{}
        \EndFor{} \\
        \STATE \hspace{\algorithmicindent} Calculate probabilities corresponding to contingency table: \\
        \hspace{\algorithmicindent} $\mathcal{P} = [$ \textcolor{red}{\texttt{\ \ \ \# p no disease vs p disease given...}} \\
        \hspace{\algorithmicindent} \hspace{\algorithmicindent} $\big(1 - \Pr(\mathcal{D})\big)$, \textcolor{red}{\texttt{\ \ \ \# no treat $\neg T$, no gene $\neg G$}} \\ %\textcolor{green}{\texttt{\# $\Pr(\neg\mathcal{D} \mid \neg G, \neg T)$}} \\
        \hspace{\algorithmicindent} \hspace{\algorithmicindent} $\Pr(\mathcal{D})$, \\
        \hspace{\algorithmicindent} \hspace{\algorithmicindent} $\big(1 - \Pr(\mathcal{D})\big) + \Pr(\mathcal{D}) \cdot \Pr(\neg \mathcal{D}|T)$, \textcolor{red}{\texttt{\ \ \ \ \ \# $T$, $\neg G$}} \\
        \hspace{\algorithmicindent} \hspace{\algorithmicindent} $\Pr(\mathcal{D}) \cdot \big(1 - \Pr(\neg \mathcal{D}|T)\big)$, \\
        \hspace{\algorithmicindent} \hspace{\algorithmicindent} $\big(1 - \Pr(\mathcal{D} \mid G)\big) \cdot \big(1 - \Pr(\mathcal{D})\big)$, \textcolor{red}{\texttt{\ \ \ \ \ \ \ \# $\neg T$, $G$}}\\
        \hspace{\algorithmicindent} \hspace{\algorithmicindent} $\Pr(\mathcal{D} \mid G) + \Pr(\mathcal{D}) - \Pr(\mathcal{D} \mid G) \cdot \Pr(\mathcal{D})$, \\
        \hspace{\algorithmicindent} \hspace{\algorithmicindent} $\big(\Pr(\mathcal{D} \mid G) + \Pr(\mathcal{D}) - \Pr(\mathcal{D} \mid G) \cdot \Pr(\mathcal{D})\big)$ \textcolor{red}{\texttt{\# $T$, $G$}} \\ \hspace{\algorithmicindent} \hspace{\algorithmicindent} \hspace{\algorithmicindent} $\cdot \Pr(\neg \mathcal{D}|T) + \big(1 - \Pr(\mathcal{D} \mid G)\big) \cdot \big(1 - \Pr(\mathcal{D})\big)$, \\
        \hspace{\algorithmicindent} \hspace{\algorithmicindent} $\big(\Pr(\mathcal{D} \mid G) + \Pr(\mathcal{D}) - \Pr(\mathcal{D} \mid G) \cdot \Pr(\mathcal{D})\big)$ \\ \hspace{\algorithmicindent} \hspace{\algorithmicindent} \hspace{\algorithmicindent} $\cdot \big(1 - \Pr(\neg \mathcal{D}|T)\big)$ \\
        \hspace{\algorithmicindent} $]$ \\
        \STATE \hspace{\algorithmicindent} \Return average log likelihood of data: \\
        \hspace{\algorithmicindent} $\mathit{log\_lik} = \sum_{i \in \{1,...,m\}}{\sum{\big(C \cdot \log(\mathcal{P})\big)}} - \log(m)$
    \EndSubAlgorithm{}
    \item \textcolor{red}{\texttt{\textbf{\# Main: causal support calculation}}} \\
    \STATE Calculate likelihood of data given causal explanations A and B: \\
    $\mathit{log\_lik}_A = \textsc{monte\_carlo}(C, 10000, \{\})$ \\
    $\mathit{log\_lik}_B = \textsc{monte\_carlo}(C, 10000, \{\Pr(\neg \mathcal{D}|T)\})$ \\
    \STATE \Return causal support for explanation A: \textcolor{red}{\texttt{\ \# Bayesian update }}\\
    $\mathfrak{cs}_A = \big(\mathit{log\_lik}_A - \mathit{log\_lik}_B\big) + \big(\log(0.5) - \log(0.5)\big)$
\end{algorithmic}
\end{algorithm}

\noindent
\textbf{Labeling ground truth causal support for each data set.}
We operationalized the ground truth for causal inferences using Griffiths and Tenenbaum's \textit{causal support}, a Bayesian cognition model that estimates the posterior log odds of a target data generating model over a set of alternative data generating models, given a data set.
In Experiment 1, we targeted causal support for explanation A over explanation B:
\begingroup
\setlength\abovedisplayskip{3pt}
\setlength\belowdisplayskip{3pt}
    \begin{flalign*}
    \setlength\abovedisplayskip{0pt}
    \setlength\belowdisplayskip{0pt}
    \mathfrak{cs}_A 
    %&= \log \bigg( \frac{\Pr(\mathit{model_A}|\mathit{data})}{\Pr(\mathit{model_B}|\mathit{data})} \bigg) \\
    &= \log \bigg( \frac{\Pr(C|\mathit{model_A})}{\Pr(C|\mathit{model_B})} \bigg) + \log \bigg( \frac{\Pr(\mathit{model_A})}{\Pr(\mathit{model_B})} \bigg)
    \end{flalign*}
\endgroup
where $C$ is the data set we label with ground truth, and models $\mathit{model_A}$ and $\mathit{model_B}$ correspond to causal explanations A and B (Fig.~\ref{fig:dags-e1-explainer}).
% Thus, we calculated causal support for explanation A ($\mathit{causal\_support_A}$) by adding together a log likelihood ratio for how strongly a data set corresponds with each generative model and a log ratio of the prior probability for each model. \jessica{We assume equal prior probability per model to simplify the communication of the prior to participants.}

The first term in the formula for $\mathfrak{cs}_A$ is a log likelihood ratio representing the relative compatibility of a given data set with causal explanations A and B. 
We computed the log likelihood of each data set 
% we showed to participants 
given $\mathit{model_A}$ and $\mathit{model_B}$ using Monte Carlo simulations (Alg.~\ref{alg:monte_carlo}, lines 28-30),
% \jessica{One thing readers maybe confused by is why we are doing this - can we motivate why we need to do it through more simulation? sort of like you provided a reason for simulation above (don't want data to appear contrived).}
% Monte Carlo simulations were 
based on structural models similar to those we used to generate data sets. 
In practical scenarios, we would not know the true data generating parameters, so we used Monte Carlo simulations of possible parameter values under each model to calculate likelihoods without needing to know the ground truth \textit{a priori}.
Under $\mathit{model_A}$ we sampled all three parameters uniformly on the interval $[0, 1]$, representing the assumptions that there is a treatment effect and that both gene and unobserved factors cause disease.
Under $\mathit{model_B}$ we sampled $\Pr(\mathcal{D}|G)$ and $\Pr(\mathcal{D})$ uniformly, but we fixed $\Pr(\neg \mathcal{D}|T)$ at zero, representing an assumption of no treatment effect (i.e., omitting the DAG arrow between treatment and gene in Fig.~\ref{fig:dags-e1-explainer}).
In each simulation, we averaged log likelihood of a given data set over $m=10000$ Monte Carlo iterations (Alg.~\ref{alg:monte_carlo}, lines 25-26), marginalizing over sampled parameter values.

The second term in the formula for $\mathfrak{cs}_A$ is a log ratio of the prior probability of explanations A versus B.
% Following Griffiths and Tenenbaum~\cite{Griffiths2005}, we assume a uniform prior across possible data generating models to be normative, such that the 
%prior
% \textit{log prior ratio} ($\mathit{lpr_A}$) 
% factors out of the formula for causal support in Experiment 1 (Alg.~\ref{alg:monte_carlo}, lines 31-32).
Following Griffiths and Tenenbaum~\cite{Griffiths2005}, we assume a uniform prior to be normative, assigning 50\% probability to both explanations A and B (Alg.~\ref{alg:monte_carlo}, lines 31-32).
% Since there were only two alternative explanations in Experiment 1, we calculated ground truth causal support using the \textit{log prior ratio} ($\mathit{lpr_A}$), %\yifan{maybe add a paren around lpr to make it more readable? also later for lrr.}
% \begingroup
% \setlength\abovedisplayskip{3pt}
% \setlength\belowdisplayskip{3pt}
%     \begin{flalign*}
%     \setlength\abovedisplayskip{0pt}
%     \setlength\belowdisplayskip{0pt}
%     \mathit{lpr_A} &= \log \bigg( \frac{\Pr(\mathit{model_A})}{\Pr(\mathit{model_B})} \bigg) = \log \bigg( \frac{0.5}{0.5} \bigg) = 0
%     \end{flalign*}
% \endgroup
% which factors out of the formula for causal support, making the likelihood itself the normative posterior for Experiment 1.
% In Experiment 2, where there are four alternative causal explanations, the prior will not factor out of the normative update in the same way.
% One can think of the prior as encoding a systematic bias in belief allocation across a finite set of alternative causal explanations.
The prior encodes a bias in belief allocation across a finite set of alternative causal explanations.
% We assume a uniform prior in defining ground truth causal support because we want our benchmark to reflect no \textit{a priori} bias towards any particular causal explanation.\footnote{Uniform priors follow a convention of psychometric models that assume guessing responses are informed by the number of response alternatives~\cite{Kingdom2010}.}
We assume a uniform prior because we want our benchmark to reflect no \textit{a priori} bias toward causal explanations.\footnote{Uniform priors follow a convention of psychometric models that assume guessing responses are informed by the number of response alternatives~\cite{Kingdom2010}.}
% \jessica{this chunk on priors makes me think we need more structure in this section. Like, define causal support as you do in the beginning, maybe then put this prior part, maybe with other things we decided related to our goals for hte datasets we create. Then we have a subsection called Dataset Generation Process or something like that which is just the order in which we did what steps.}

\subsubsection{Performance evaluation}
We wanted to measure how much participants' causal inferences deviated from our normative benchmark, causal support. 

% \noindent
% \textbf{Perceived causal support.}
% We compare participants' responses to causal support on a log odds scale.
% This requires transforming responses into \textit{perceived causal support}, a log response ratio ($\mathit{lrr_A}$),
% \begingroup
% \setlength\abovedisplayskip{3pt}
% \setlength\belowdisplayskip{3pt}
%     $$\mathit{lrr_A} = \log \bigg( \frac{\mathit{response_A}}{\mathit{response_B}} \bigg)$$
% \endgroup
% where $\mathit{response_A}$ and $\mathit{response_B}$ were the probabilities participants allocated to causal explanations A and B, respectively, on each trial.

\noindent
\textbf{The linear in log odds model \& causal support.}
By choosing to model perceived causal support $\mathit{lrr_A}$ (see Section 3.1.1) as a function of ground truth causal support on a log odds scale, we leverage a linear in log odds (LLO) model to extend causal support from a normative cognitive model into a descriptive one.
Prior work shows that the LLO model accurately describes natural distortions in mental representations of probability~\cite{Gonzalez1999,Hollands2000,Stevens1957,Zhang2012}.
% For example, Zhang et al.~\cite{Zhang2012} apply the LLO model to reanalyze data from multiple experiments involving a variety of judgments of probabilities and proportions, finding that the functional relationship between actual and perceived proportions seems to be ubiquitously LLO.
% Related literature in visualization~\cite{Kale2021} uses the LLO model to measure perceptual distortions in probabilistic judgments about intervention effectiveness.
For example, visualization researchers~\cite{Kale2021} used the LLO model to measure perceptual distortions in probabilistic judgments about intervention effectiveness.
Our normative model of causal support itself (see Section 3.1.4) is 
%a LLO model, a sum of likelihoods and priors in log odds units.
a sum in log odds units.

\begin{figure*}[t]
    \centering
    \includegraphics[width=\textwidth]{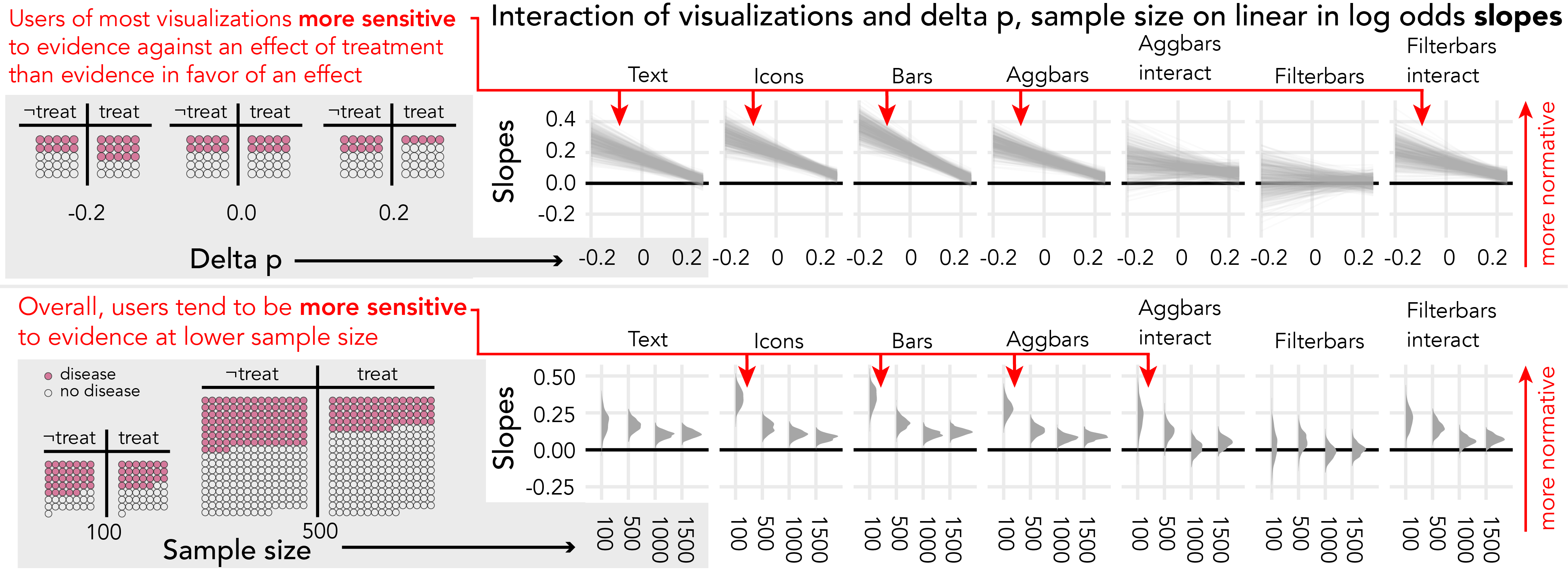}
    \setlength{\abovecaptionskip}{-10pt}
    \setlength{\belowcaptionskip}{-17pt}
    \caption{Sensitivity (y-axes) conditioned on two attributes of visual signal for treatment effectiveness (rows, x-axes) and visualizations (columns).
    }
    \label{fig:e1-interactions}
\end{figure*}

\begin{figure}[t]
    \centering
    \includegraphics[width=\columnwidth]{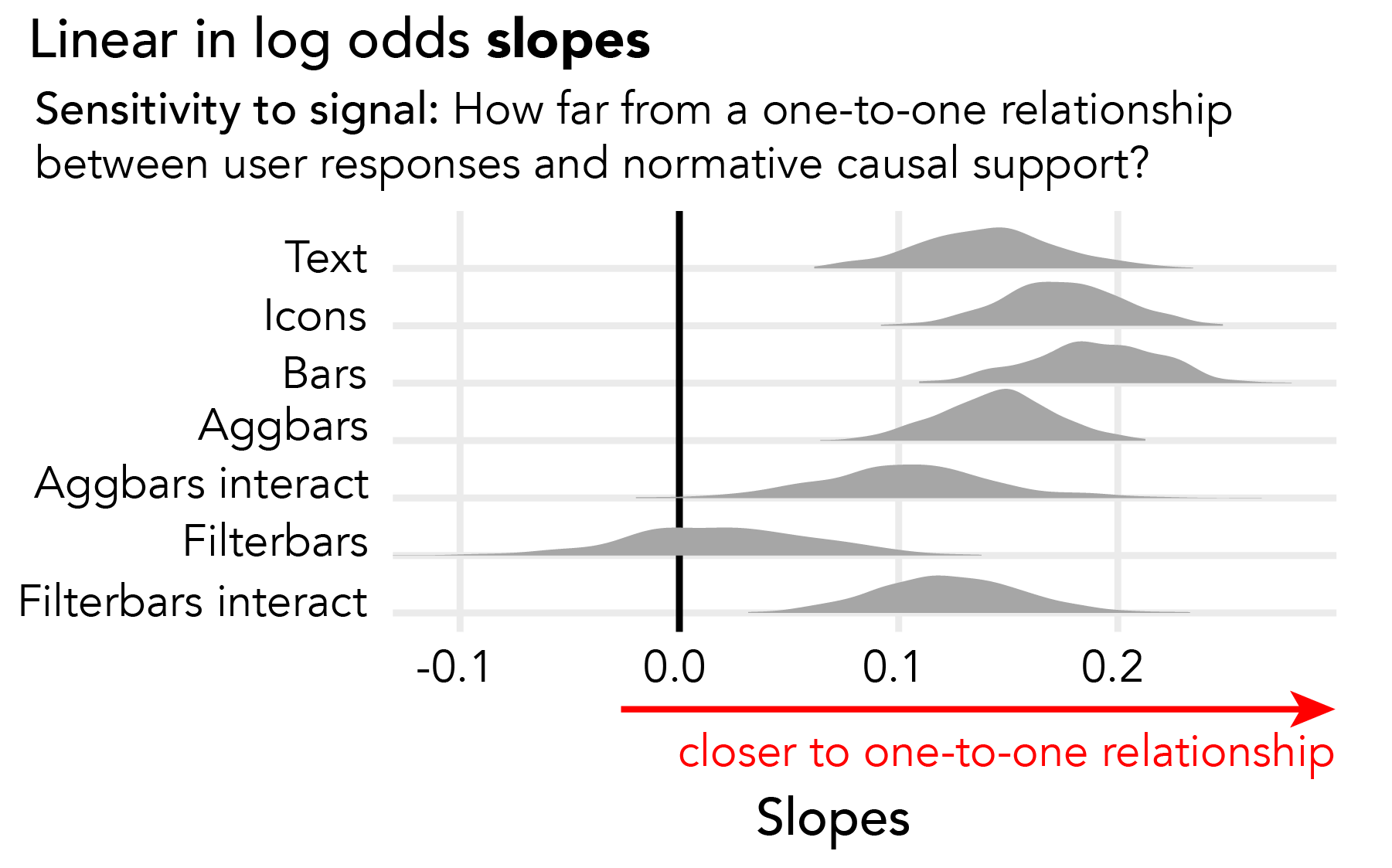}
    \setlength{\abovecaptionskip}{-10pt}
    \setlength{\belowcaptionskip}{0pt}
    \caption{Linear in log odds (LLO) slopes per visualization condition.
    }
    \label{fig:e1-slopes}
\end{figure}

\noindent
\textbf{Derived measures.}
Using a LLO model to measure the correspondence between normative and perceived causal support enables us to estimate (1) participants' \textit{sensitivity} to changes in ground truth causal support and (2) \textit{bias} in perceived causal support.
From our model, we derive sensitivity and bias per condition as \textit{LLO slopes and intercepts}, respectively.
LLO slopes describe sensitivity to ground truth causal support such that a slope of one indicates ideal sensitivity.
One can think of slopes as the weight participants assign to changes in the ground truth log likelihood ratio of explanations A versus B.
LLO intercepts describe bias in participants' probability allocations when causal support is zero such that an intercept of 50\% indicates no bias.
One can think of intercepts as the average prior probability that participants allocate to explanation A when there is no signal in the data.

\noindent
\textbf{Approach.}
We used the brms package~\cite{Burkner2020} in R to fit Bayesian hierarchical models on perceived causal support.
We adopted a Bayesian workflow called \textit{model expansion}~\cite{Gabry2019}, where we started with a simple model and iteratively added predictors to build up to more complex models, running prior predictive checks, model diagnostics, posterior predictive checks, and leave-one-out cross validation for each version of the model.
We centered each prior to reflect a null hypothesis of ideal performance and no bias, and we scaled each prior to be weakly informative while providing sufficient regularization for models to converge.
We provide more details about our modeling workflow in our preregistrations\footnote{
% We include an anonymized preregistration in Supplemental Materials for review. There is a version with author names on OSF. 
See preregistrations for Experiment 1 (\url{https://osf.io/vzmhu}) and for Experiment 2 (\url{https://osf.io/y46nw})
} and Supplemental Materials.\footnote{
\url{https://github.com/kalealex/causal-support}
% \texttt{Github repo omitted for anonymous review.}
}

\noindent
\textbf{Model specification.}
We used the following model (Wilkinson-Pinheiro-Bates notation~\cite{Wilkinson1973,Burkner2020,Pinheiro2020}) to evaluate participants' responses: 
\begingroup
\setlength\abovedisplayskip{3pt}
\setlength\belowdisplayskip{3pt}
    \begin{flalign*}
    \setlength\abovedisplayskip{0pt}
    \setlength\belowdisplayskip{0pt}
    \mathit{lrr_A} \sim &\mathrm{Normal}(\mu_A, \sigma_A) \\[-2pt]
    \mu_A = &\mathfrak{cs}_A*\mathit{delta\_p}*n*\mathit{vis} %\\[-2pt]
    + \big(\mathfrak{cs}_A*\mathit{delta\_p} + \mathfrak{cs}_A*n\big|\mathit{worker\_id}\big)
    \end{flalign*}
\endgroup
where $\mathit{lrr_A}$ was perceived causal support for a treatment effect,
% based on participant responses, 
$\mathfrak{cs}_A$ was our normative benchmark, $\mathit{delta\_p}$ was the difference in the proportion of people with disease given treatment versus no treatment, $n$ was the sample size as a factor, $\mathit{vis}$ was a dummy variable for visualization condition, and $\mathit{worker\_id}$ was a unique identifier for each participant.

We primarily modeled effects on the mean of perceived causal support $\mu_A$, but our model also learned the residual standard deviation $\sigma_A$.
% of perceived causal support.
Both $\sigma_A$ and the random effects in the $\mu_A$ submodel helped account for the empirical distribution, differentiating between response noise and effects of interest.
The term $\mathfrak{cs}_A*\mathit{delta\_p}*n*\mathit{vis}$ enabled our model to learn how the slope on causal support varies as a function of the visual signal on each trial
($\mathit{delta\_p}$ and $n$)
% ---operationalized as $\mathit{delta\_p}$ and $n$---as well as
and visualization condition.
% (i.e., all of these factors interacted with each other).

\subsubsection{Participants \& exclusions}
We recruited participants on Amazon Mechanical Turk.
Workers had a HIT acceptance rate of at least 97\% and were located in the US.
We aimed to recruit a total of 400 participants after exclusions using our attention check trials, 80 per visualization condition.
We determined this target sample size using a heuristic power analysis based on pilot data and the assumption that the width of confidence intervals would be inversely proportional to $\sqrt{N}$.
We recruited a total of 548 participants, and after exclusions we used data from 408 participants in our analysis. 
We slightly overshot our target sample size because we could not anticipate perfectly how many participants would miss our attention checks (see Section 3.1.4).
Although we preregistered that we would exclude participants who failed to allocate at least 50\% subjective probability to the most likely causal explanation on either attention check, this criterion proved too strict and would have excluded 48\% of our sample.
Instead, we opted to use only the easier of the two attention checks for exclusions, resulting in the exclusion of 26\% of our sample.
All participants were paid regardless of exclusions.
We compensated the average participant \$2.50 for about 9 minutes.
\subsection{Results}
% We compare chart users' causal inferences to a benchmark called causal support, \jessica{seems awkward since we've already defined it} using a linear in log odds (LLO) model to describe performance in terms of \textit{sensitivity} to the ground truth and \textit{bias} in probability allocations when each causal explanation is equally likely.
We evaluate chart users' causal inferences using a linear in log odds (LLO) model to assess \textit{sensitivity} to the ground truth and \textit{bias} in probability allocations when each causal explanation is equally likely.

\noindent
\textbf{Sensitivity.}
A LLO slope of one indicates one-to-one correspondence between the ground truth and users' probability allocations.
In all visualization conditions (Fig.~\ref{fig:e1-slopes}), we see slopes far below one, indicating that users are much less sensitive than ideal.
% On average, t
The only reliable differences between visualization conditions are that filterbars users who do not interact are less sensitive than users in other conditions. 

When filterbars users do not interact with the charts, slopes are approximately zero indicating that users are insensitive to signal.
Performance improves reliably when users interact with the visualization by applying cross-filters to coordinated multiple views.
This is expected because filterbars hide visual signal for the task behind interactions.

Surprisingly, when aggbars users interact with the charts to group by gene or treatment, this leads to lower sensitivity, though this difference is not reliable. 
To make sense of this result, we analyze interaction log data to see which variables chart users condition on.
We find that aggbars users group the data by gene almost as often as treatment.
Compare this to filterbars users, who condition on treatment much more often than gene (see Supplemental Material).
This suggests that interacting with visualizations only improves sensitivity to causal support when users deliberately generate views of the data which show counterfactual predictions that can distinguish competing causal explanations. 

\noindent
\textbf{Visual signal effects on sensitivity.} 
We examine sensitivity in each visualization as function of attributes of the visual signal for our task.
In Experiment 1, the signal breaks down into two data attributes, \textit{delta p} and \textit{sample size} (see Section 3.1.4).
Normatively, LLO slopes equal one regardless of delta p and sample size, however, our model measures differences in sensitivity depending on these data attributes.

Figure~\ref{fig:e1-interactions} shows that in the conditions where slopes are largest---text, icons, bars, aggbars without interaction, and filterbars with interaction---users are more sensitive to causal support at negative values of delta p (e.g., Fig.~\ref{fig:e1-interactions}, top inset).
The average user in these conditions responds more to evidence against treatment (i.e., falsification) than evidence in favor of a treatment effect (i.e., verification). 
At positive values of delta p (e.g., Fig.~\ref{fig:e1-interactions}, top inset), LLO slopes are similar across conditions, suggesting that differences in performance between conditions are driven in part by differences in sensitivity to falsifying evidence.

We also see in Figure~\ref{fig:e1-interactions} that users of icons, bars, and aggbars 
% especially 
are more sensitive to signal when sample size is smaller. 
This finding is consistent with prior work showing that chart users tend to underestimate sample size when making inferences with data~\cite{Kim2019}, which may be driven by logarithmic perception~\cite{Varshney2013,Zhang2012}.
Alternatively, we could interpret this result as a cognitive bias where users are unwilling to be certain even when sample size is large enough to support unambiguous inferences, related to non-belief in the law of large numbers~\cite{Benjamin2016}.
% Whatever the reason, we find it somewhat surprising that insensitivity to sample size persists even with icon arrays, which emphasize sample size as the overall number of equal-sized dots.

\begin{figure}[t]
    \centering
    \includegraphics[width=\columnwidth]{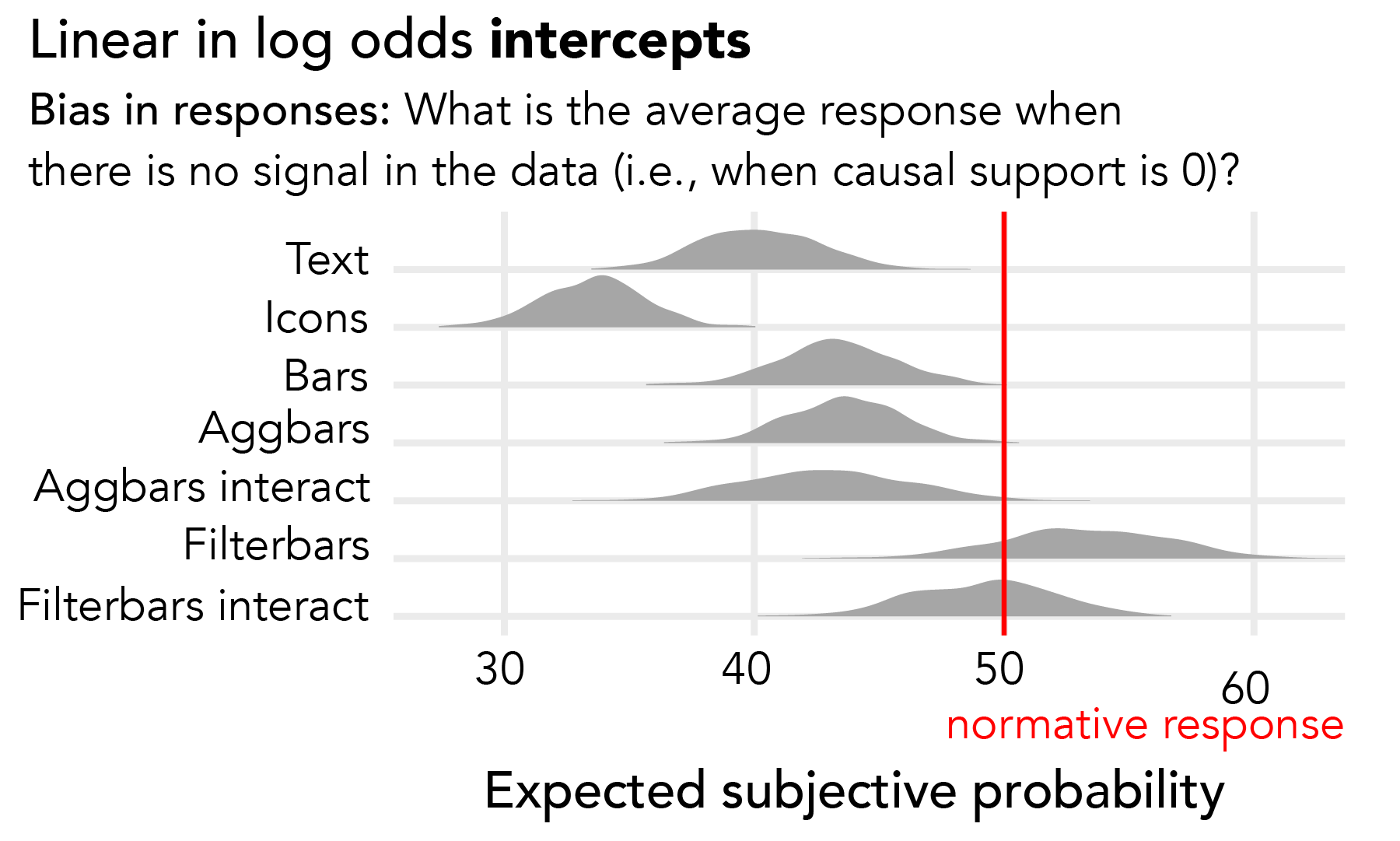}
    \setlength{\abovecaptionskip}{-10pt}
    \setlength{\belowcaptionskip}{0pt}
    \caption{LLO intercepts per visualization condition.
    }
    \label{fig:e1-intercepts}
\end{figure}

\noindent
\textbf{Bias.}
Intercepts in the LLO model describe bias in users' probability allocations when ground truth causal support indicates that explanation A (i.e., treatment effect) is just as likely as explanation B (i.e., no treatment effect). 
Under this condition, a normative observer would allocate equal probability to both causal explanations.
We derive expected probability allocated to explanation A based on a logistic transform of LLO intercepts, and compare this to the normative benchmark of 50\%.

With all visualizations except for filterbars, probability allocations are far below 50\% indicating substantial bias (Fig.~\ref{fig:e1-intercepts}). 
On average when causal support is zero, users of text tables, icons, bars, and aggbars allocate too little probability to causal explanation A. 
Users of filterbars, on the other hand, allocate approximately 50\% to explanation A.
We see the most extreme bias of up to 20\% with icons arrays.

Unfortunately, we can only speculate about possible reasons for these biases.
We expected that LLO intercepts would indicate average responses near 50\% in the absence of signal for all conditions (i.e., a uniform prior), simply because this follows from the structure of the task. 
Because this pattern of biases across visualizations results from a non-preregistered exploratory comparison, we investigate in Experiment 2 whether these biases replicate for a more complex task.
\section{Experiment 2}
In Experiment 2, we evaluate the same visualization designs on a more difficult task. 
We asked participants to detect confounding in the presence of a known treatment effect by allocating probability across four possible ``backdoor paths''~\cite{Pearl2018} (Fig.~\ref{fig:dags-e2-explainer}). %\jessica{have we explained backdoor path? if not cite something and put in ""}
We extend causal support to handle more than two alternative causal explanations, demonstrating how causal support can be employed in more complex analyses.

\subsection{Method}
Experiment 2 was the same as Experiment 1 except for the following changes to response elicitation, modeling, and experimental design.

\subsubsection{Task scenario \& response elicitation}
\begin{figure}[t]
    \centering
    \includegraphics[width=\columnwidth]{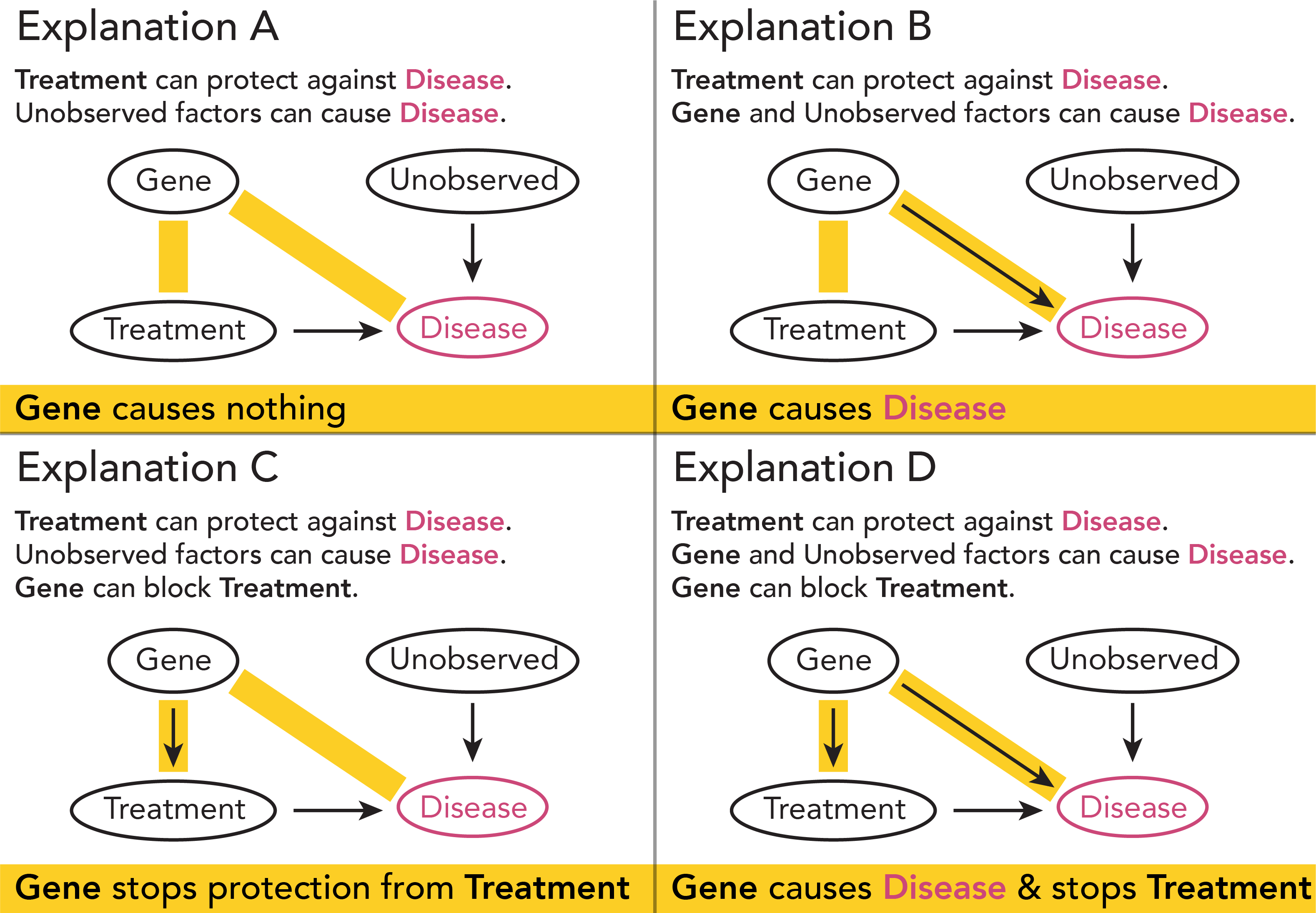}
    \setlength{\abovecaptionskip}{-10pt}
    \setlength{\belowcaptionskip}{0pt}
    \caption{DAGs for possible causal explanations in Experiment 2.
    }
    \label{fig:dags-e2-explainer}
\end{figure}

% We asked participants to judge the influence of a gene on both disease and treatment effectiveness by allocating probability across the four DAGs in Figure~\ref{fig:dags-e2-explainer}.
% Participants separately judge each component effect of a potential confounding relationship.
Participants judge the influence of a gene on both disease and treatment effectiveness by allocating probability across the four DAGs in Figure~\ref{fig:dags-e2-explainer}, separately assessing each DAG arrow in a confounding relationship.

\noindent
\textbf{Question \& Elicitation.}
We asked participants a similar question as in Experiment 1, where participants allocated 100 votes (i.e., subjective probability) across alternative causal explanations.
However, in Experiment 2 we elicited a Dirichlet distribution with four alternatives. 
Following Chalone et al.~\cite{Chalone1987,OHagan2006} and extending our interface from Experiment 1, 
each time participants allocated a number of votes between 0 and 100 to an option, the \textit{remaining votes out of 100 were uniformly distributed across unused response options}. 
These imputed responses were highlighted along with a prompt to, \textit{``Adjust your responses until all the numbers reflect your beliefs.''}
Participants iteratively set and adjusted their probability allocations. 
% We combined the four resulting probabilities ($\mathit{response_A}$, $\mathit{response_B}$, $\mathit{response_C}$, and $\mathit{response_D}$) into estimates of perceived causal support, which we compared to our normative benchmark. 
We combined these responses into perceived causal support, which we compared to our benchmark. 

\noindent
\textbf{Perceived causal support.}
When estimating \textit{perceived causal support} in Experiment 2, we separately evaluated \textit{multiple target explanations}.
Primarily, we targeted belief in explanation D ($\mathit{llr_{D}}$), confounding:
\begingroup
\setlength\abovedisplayskip{3pt}
\setlength\belowdisplayskip{3pt}
    \begin{flalign*}
    \setlength\abovedisplayskip{0pt}
    \setlength\belowdisplayskip{0pt}
    \mathit{lrr_D} &= \log \bigg( \frac{\mathit{response_D}}{\sum_{i=\{A,B,C\}}\mathit{response_i}} \bigg)
    \end{flalign*}
\endgroup
where $\mathit{response_A}$, $\mathit{response_B}$, $\mathit{response_C}$, and $\mathit{response_D}$ were participants' probability allocations to causal explanations A through D, respectively, on each trial.
We also separately targeted belief in both of the component DAG arrows that constitute a confounding relationship (Fig.~\ref{fig:dags-e2-explainer}): ($\mathit{llr_{BD}}$) the effect of gene on disease, which appears in explanations B and D; and ($\mathit{llr_{CD}}$) the effect of gene on treatment effectiveness, which appears in explanations C and D. We define $\mathit{llr_{BD}}$ as follows,
\begingroup
\setlength\abovedisplayskip{3pt}
\setlength\belowdisplayskip{3pt}
    \begin{flalign*}
    \setlength\abovedisplayskip{0pt}
    \setlength\belowdisplayskip{0pt}
    \mathit{lrr_{BD}} &= \log \bigg( \frac{\sum_{i=\{B,D\}}\mathit{response_i}}{\sum_{i=\{A,C\}}\mathit{response_i}} \bigg) %\\
    % lrr_{CD} &= \log \bigg( \frac{\sum_{i=\{C,D\}}response_i / 100}{\sum_{i=\{A,B\}}response_i / 100} \bigg)
    \end{flalign*}
\endgroup
and we define $\mathit{llr_{CD}}$ similarly. 
We compare log response ratios $\mathit{llr_{D}}$, $\mathit{llr_{BD}}$, and $\mathit{llr_{CD}}$ to corresponding causal support $\mathfrak{cs}_D$, $\mathfrak{cs}_\mathit{BD}$ and $\mathfrak{cs}_\mathit{CD}$.

% \noindent
% \textbf{Payment.}
% In Experiment 2, participants received a \$0.25 bonus for every trial where they responded within 5 percentage points of the ground truth probability of causal explanation D (Fig.~\ref{fig:dags-e2-explainer}).

\noindent
\textbf{Strategy.}
At the end of the experiment, we asked participants,
\begin{quote}
    \vspace{-3pt}
    \textit{\textbf{How did you use the charts to complete the task?} Please tell us what patterns you looked for in the data and what comparisons you made.}
\end{quote}
\vspace{-3pt}
\noindent
We analyzed these qualitative responses to assess whether participants understood how to use the charts for the confounding detection task.

\subsubsection{Experimental design}
% We manipulated both the visualizations participants used for the task (between-subjects) and the data sets that we showed participants.
We manipulated both the visualizations (between-subjects) and the data sets we showed (within-subjects).
% We showed each participant a total of 19 trials, 18 data conditions (see Section 4.1.3) presented within-subjects and one attention check used for exclusions (see Section 4.1.5). We randomized trial order for each participant, inserting the attention check on trial 10.
We showed each participant 19 trials, 18 data conditions (see Section 4.1.3) and one attention check used for exclusions (see Section 4.1.5). We randomized trial order for each participant, inserting the attention check on trial 10.

\subsubsection{Stimulus generation}
We generated data sets that spanned a range of ground truth causal support \textit{for confounding}.
Creating these data sets required (1) manipulating data attributes which signaled whether the gene was a confounding factor, and (2) labeling each data set with ground truth causal support.

Our goal was to generate 18 data conditions that varied delta p disease, delta p treatment, and sample size, data attributes which manipulate causal support for confounding.
\textit{Delta p disease} described the difference in the proportion of people with disease in each data set depending on whether they had the gene.
Negative values of delta p disease indicated evidence that the gene caused disease, whereas values near zero indicated evidence against a gene effect on disease.
\textit{Delta p treatment} described the difference in the proportion of people with disease within the treatment group depending on whether they had the gene.
Negative values of delta p treatment indicated evidence that the gene stopped the treatment from preventing disease, whereas values near zero indicated evidence against a gene effect on treatment.
\textit{Sample size} was the number of people in each data set we showed chart users.

\noindent
\textbf{Data conditions.}
To generate 18 data conditions, we simulated data from structural models with one parameter per DAG arrow in Figure~\ref{fig:dags-e2-explainer}.
We manipulated the \textit{probability that gene causes disease} (3 levels: $\{0, 0.35, 0.7\}$), the \textit{probability that gene prevents treatment from working} (3 levels: $\{0, 0.35, 0.7\}$), and \textit{sample size} (2 levels: $\{100, 1000\}$).
We controlled the probability that treatment prevents disease ($0.8$), probability of disease due to unobserved causes ($0.2$), base rate of the gene ($0.4$), and the proportion of each sample with treatment ($0.5$).
We selected these parameters iteratively by sampling data sets and labeling ground truth until half of trials had greater than a 25\% chance of having been generated by causal explanation D.

As in Experiment 1, we simulated many data sets for each data condition, and we counterbalanced quantiles of sampling error across participants (see Section 3.1.4).
For our attention check trial, we selected the simulated data set that maximized causal support for confounding.
% These within-subjects manipulations result in 18 sets of realistic-looking count data per participant.

% We showed participants one attention check trial, chosen to maximize causal support for confounding. We presented this attention check on the 10th of 19 total trials and used it for our exclusion criterion.

\noindent
\textbf{Labeling ground truth causal support.}
We extended Griffiths and Tenenbaum's model of causal support~\cite{Griffiths2005} to account for more than two alternative causal explanations. 
% We were interested primarily in causal support for causal explanation D, 
We primarily targeted causal support for causal explanation D over explanations A, B or, C, 
\begingroup
\setlength\abovedisplayskip{3pt}
\setlength\belowdisplayskip{3pt}
    \begin{flalign*}
    \setlength\abovedisplayskip{0pt}
    \setlength\belowdisplayskip{0pt}
    \mathfrak{cs}_D %=& \log \bigg( \frac{\Pr(\mathit{model_D}|\mathit{data})}{\sum_{i=\{A,B,C\}}\Pr(\mathit{model_i}|\mathit{data})} \bigg) \\
    =& \log \bigg( \frac{\Pr(C|\mathit{model_D})}{\sum_{i=\{A,B,C\}}\Pr(C|\mathit{model_i})} \bigg) %\\ 
    + \log \bigg( \frac{\Pr(\mathit{model_D})}{\sum_{i=\{A,B,C\}}\Pr(\mathit{model_i})} \bigg)
    \end{flalign*}
\endgroup
where $C$ is the data set we label with ground truth, and $\mathit{model_A}$, $\mathit{model_B}$, $\mathit{model_C}$, and $\mathit{model_D}$ correspond to causal explanations A through D (Fig.~\ref{fig:dags-e2-explainer}), respectively.
Since we separately targeted belief in both of the component DAG arrows that constitute a confounding relationship (see Section 4.1.1, perceived causal support), we needed to calculate ($\mathfrak{cs}_\mathit{BD}$) ground truth causal support for explanations B or D over A or C:
% Detecting confounding requires interpreting two distinct effects, the effect of gene on disease (Fig.~\ref{fig:dags-e2-explainer}, explanation B or D) and the effect of gene on treatment effectiveness (Fig.~\ref{fig:dags-e2-explainer}, explanation C or D).
% To study participants' sensitivity to each of these effects separately, we compared their responses to causal support for explanations B or D and causal support for explanations C or D, respectively\footnote{We define $\mathfrak{cs}_\mathit{BD}$ and $\mathfrak{cs}_\mathit{CD}$ similarly.}.
\begingroup
\setlength\abovedisplayskip{3pt}
\setlength\belowdisplayskip{3pt}
    \begin{flalign*}
    \setlength\abovedisplayskip{0pt}
    \setlength\belowdisplayskip{0pt}
    \mathfrak{cs}_\mathit{BD} %=& \log \bigg( \frac{\sum_{i=\{B,D\}}\Pr(model_i|D)}{\sum_{i=\{A,C\}}\Pr(model_i|D)} \bigg) \\
    =& \log \bigg( \frac{\sum_{i=\{B,D\}}\Pr(C|model_i)}{\sum_{i=\{A,C\}}\Pr(C|model_i)} \bigg) %\\ 
    + \log \bigg( \frac{\sum_{i=\{B,D\}}\Pr(model_i)}{\sum_{i=\{A,C\}}\Pr(model_i)} \bigg) 
    \end{flalign*}
\endgroup
% \begingroup
% \setlength\abovedisplayskip{3pt}
% \setlength\belowdisplayskip{3pt}
%     \begin{flalign*}
%     \setlength\abovedisplayskip{0pt}
%     \setlength\belowdisplayskip{0pt}
%     causal\_support_{CD} =& \log \bigg( \frac{\sum_{i=\{C,D\}}\Pr(model_i|D)}{\sum_{i=\{A,B\}}\Pr(model_i|D)} \bigg) \\
%     =& \log \bigg( \frac{\sum_{i=\{C,D\}}\Pr(D|model_i)}{\sum_{i=\{A,B\}}\Pr(D|model_i)} \bigg) \\ 
%     &+ \log \bigg( \frac{\sum_{i=\{C,D\}}\Pr(model_i)}{\sum_{i=\{A,B\}}\Pr(model_i)} \bigg)
%     \end{flalign*}
% \endgroup
We similarly calculated ($\mathfrak{cs}_\mathit{CD}$) causal support for explanations C or D.

% \alex{Frame as modifications to Algorithm 1.}
The first terms in the formulae for $\mathfrak{cs}_D$, $\mathfrak{cs}_{BD}$, and $\mathfrak{cs}_{CD}$ are log likelihood ratios representing the relative compatibility of a given data set with causal explanations A, B, C, and D.
We calculated the log likelihood of each data set we showed participants given $\mathit{model_A}$, $\mathit{model_B}$, $\mathit{model_C}$, and $\mathit{model_D}$ using Monte Carlo simulations similar to Algorithm~\ref{alg:monte_carlo}.
In Experiment 2, we introduced one more parameter $\Pr(\neg T|G)$ to our structural models, representing the probability that the gene prevents the treatment effect.
We incorporate this parameter into our Monte Carlo simulations (Alg.~\ref{alg:monte_carlo}) by making the following substitutions:
\begingroup
\setlength\abovedisplayskip{3pt}
\setlength\belowdisplayskip{3pt}
\begin{flalign*}
\setlength\abovedisplayskip{0pt}
\setlength\belowdisplayskip{0pt}
    \mathrm{20}: &\ (\Pr(\mathcal{D}|G) + \Pr(\mathcal{D}) - \Pr(\mathcal{D}|G) \cdot \Pr(\mathcal{D})) \cdot (\Pr(\neg \mathcal{D}|T) \\ 
    \mathrm{21}: &\ \ \ \ \cdot (1 - \Pr(\neg T|G)) + (1 - \Pr(\mathcal{D}|G)) \cdot (1 - \Pr(\mathcal{D})), \\
    \mathrm{22:} &\ (\Pr(\mathcal{D}|G) + \Pr(\mathcal{D}) - \Pr(\mathcal{D}|G) \cdot \Pr(\mathcal{D})) \\
    \mathrm{23}: &\ \ \ \ \cdot ((1 - \Pr(\neg \mathcal{D}|T)) + \Pr(\neg \mathcal{D}|T) \cdot \Pr(\neg T|G))
\end{flalign*}
\endgroup
Under $model_A$ we sampled $\Pr(\mathcal{D})$ and $\Pr(\neg \mathcal{D}|T)$ uniformly on the interval $[0, 1]$ and fixed $\Pr(\mathcal{D}|G)$ and $\Pr(\neg T|G)$ at zero, representing assumptions that the gene impacts neither disease or treatment. 
Under $model_B$ we sampled $\Pr(\mathcal{D})$, $\Pr(\mathcal{D}|G)$, and $\Pr(\neg \mathcal{D}|T)$ uniformly and fixed $\Pr(\neg T|G)$ at zero, representing the assumption that the gene has no effect on treatment. 
Under $model_C$ we sampled $\Pr(\mathcal{D})$, $\Pr(\neg T|G)$, and $\Pr(\neg \mathcal{D}|T)$ uniformly and fixed $\Pr(\mathcal{D}|G)$ at zero, representing the assumption that the gene has no effect on disease.
Under $model_D$ we sampled all four parameters uniformly to represent confounding.
The second terms in the formulae for $\mathfrak{cs}_D$, $\mathfrak{cs}_{BD}$, and $\mathfrak{cs}_{CD}$ are log ratios of the prior probabilities of the target explanation(s) versus other possible explanations.
Again, we assumed a uniform prior to create an unbiased benchmark for our task such that 25\% was the normative prior probability for each causal explanation A, B, C, and D, respectively.

\begin{figure*}[t]
    \centering
    \includegraphics[width=\textwidth]{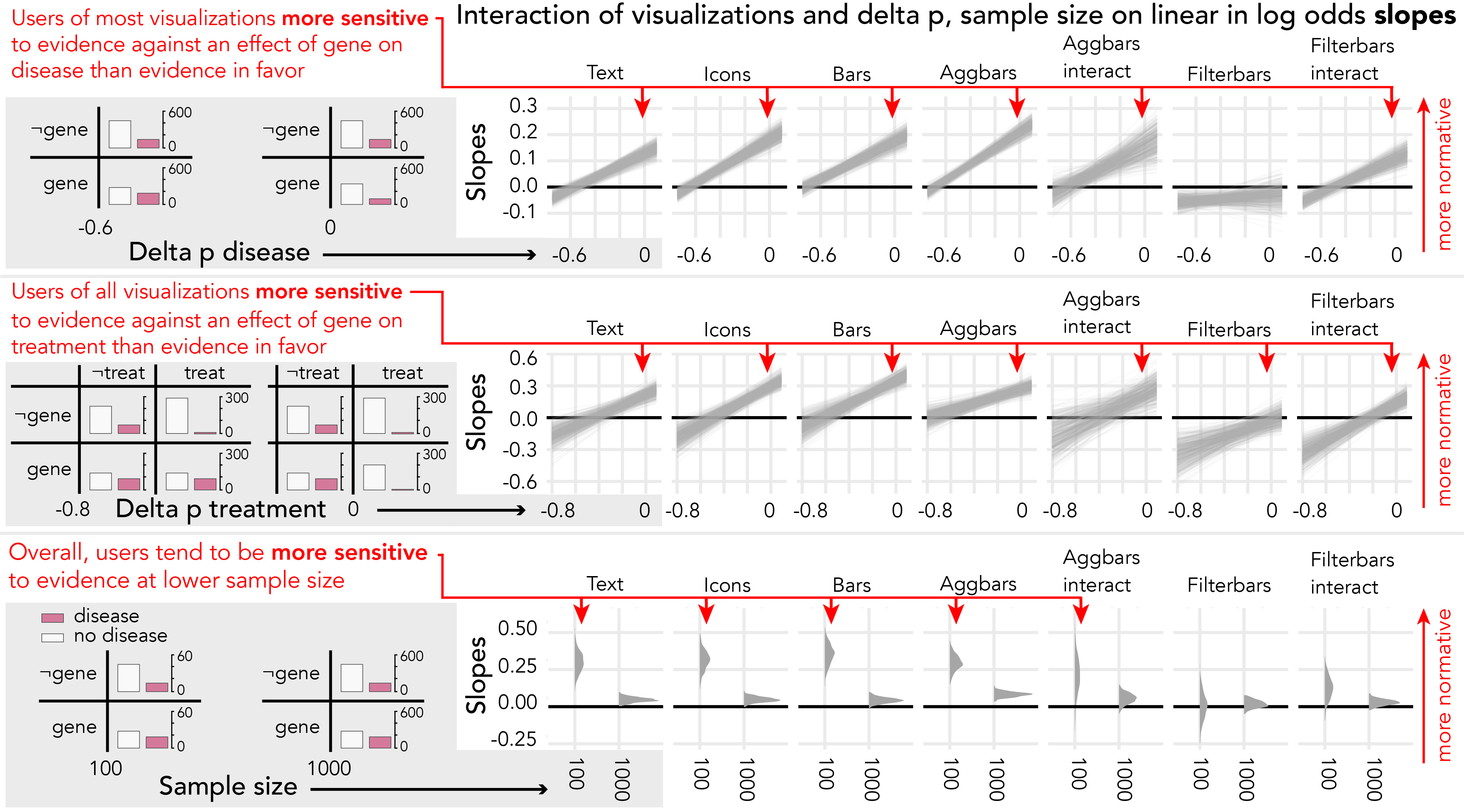}
    \setlength{\abovecaptionskip}{-10pt}
    \setlength{\belowcaptionskip}{-20pt}
    \caption{Sensitivity (y-axes) conditioned on three attributes of visual signal for confounding (rows, x-axes) and visualization conditions (columns).
    }
    \label{fig:e2-interactions}
\end{figure*}

\begin{figure}[t]
    \centering
    \includegraphics[width=\columnwidth]{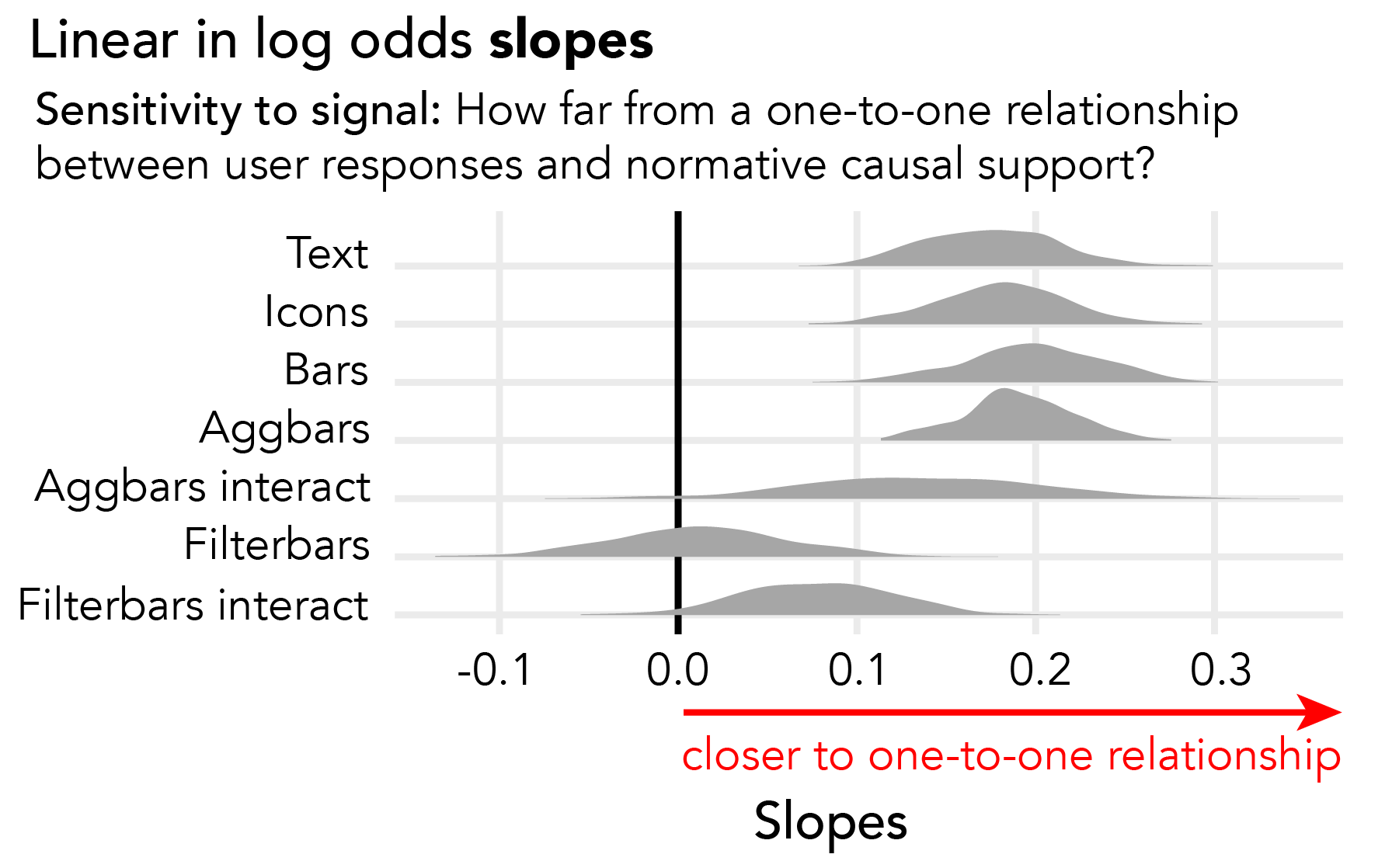}
    \setlength{\abovecaptionskip}{-10pt}
    \setlength{\belowcaptionskip}{0pt}
    \caption{Linear in log odds (LLO) slopes per visualization condition.
    }
    \label{fig:e2-slopes}
\end{figure}

\begin{figure}[t]
    \centering
    \includegraphics[width=\columnwidth]{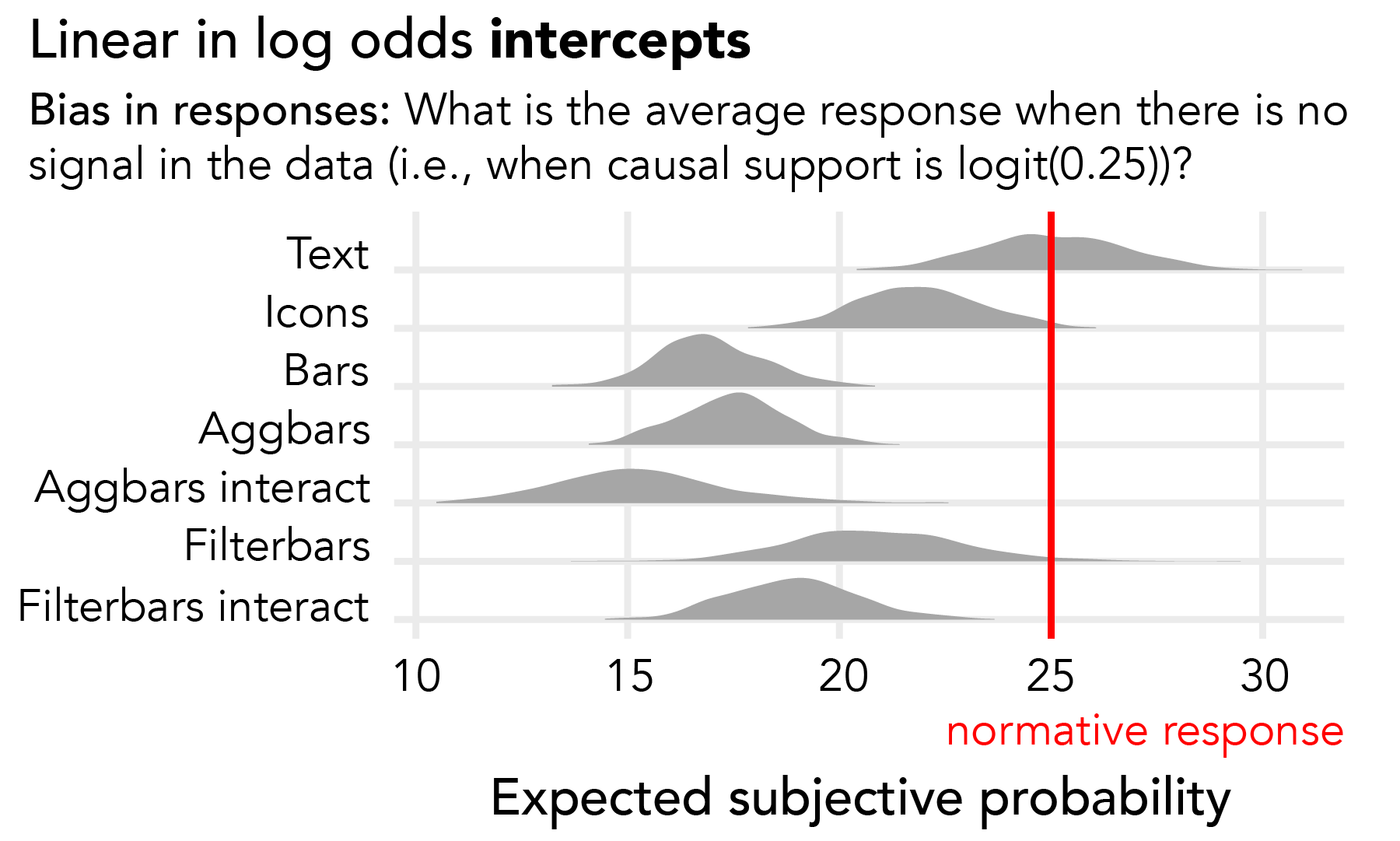}
    \setlength{\abovecaptionskip}{-10pt}
    \setlength{\belowcaptionskip}{0pt}
    \caption{LLO intercepts per visualization condition.
    }
    \label{fig:e2-intercepts}
\end{figure}

\subsubsection{Performance evaluation}
Again, we used linear in log odds (LLO) models~\cite{Gonzalez1999,Zhang2012} to describe discrepancies between perceived and normative causal support.
We also conducted a qualitative analysis of participants' reported strategies.

\noindent
\textbf{Model specification.}
% We needed three inferential models because we had three dependent variables, representing perceived causal support for confounding ($\mathit{lrr_D}$) and perceived causal support for the two constituent effects of confounding ($\mathit{lrr_{BD}}$ and $\mathit{lrr_{CD}}$).
We used three inferential models because we had three dependent variables, representing perceived causal support for confounding ($\mathit{lrr_D}$) and for the two constituent effects of confounding ($\mathit{lrr_{BD}}$ and $\mathit{lrr_{CD}}$).
Here, we show only the models on $\mathit{lrr_D}$ and  $\mathit{lrr_{BD}}$ because the models on $\mathit{lrr_{CD}}$ and $\mathit{lrr_{BD}}$ are identical in form, with $\mathfrak{cs}_\mathit{CD}$ and $\mathit{delta\_p\_t}$ replacing $\mathfrak{cs}_\mathit{BD}$ and $\mathit{delta\_p\_d}$ as predictors:
% We used these models\footnote{The models on $\mathit{lrr_{CD}}$ and $\mathit{lrr_{BD}}$ are identical in form, with $\mathit{causal\_support_{CD}}$ and $\mathit{delta\_p\_t}$ replacing $\mathit{causal\_support_{BD}}$ and $\mathit{delta\_p\_d}$ as predictors.}:
% a primary model on causal support for confounding and two secondary models examining causal support for the gene effect on disease and treatment effectiveness, respectively: 
\begingroup
\setlength\abovedisplayskip{3pt}
\setlength\belowdisplayskip{3pt}
    \begin{flalign*}
    \setlength\abovedisplayskip{0pt}
    \setlength\belowdisplayskip{0pt}
    \mathit{lrr_D} \sim &\mathrm{Normal}(\mu_D, \sigma_D) \\[-2pt]
    \mu_D = &\mathfrak{cs}_D*\mathit{delta\_p\_d}*\mathit{delta\_p\_t}*n*\mathit{vis} \\[-2pt]
    &+ \big(\mathfrak{cs}_D*\mathit{delta\_p\_d} %\\[-2pt]
    + \mathfrak{cs}_D*\mathit{delta\_p\_t} %\\[-2pt]
    + \mathfrak{cs}_D*n\big|\mathit{worker\_id}\big) \\
    \mathit{lrr_{BD}} \sim &\mathrm{Normal}(\mathit{\mu_{BD}}, \mathit{\sigma_{BD}}) \\[-2pt]
    \mathit{\mu_{BD}} = &\mathfrak{cs}_\mathit{BD}*\mathit{delta\_p\_d}*n*\mathit{vis} %\\[-2pt]
    + \big(\mathfrak{cs}_\mathit{BD}\big|\mathit{worker\_id}\big) %\\
    % lrr_{CD} \sim &\mathrm{Normal}(\mu_{CD}, \sigma_{CD}) \\[-2pt]
    % \mu_{CD} = &causal\_support_{CD}*delta\_p\_t*n*vis \\[-2pt]
    % &+ \big(causal\_support_{CD}\big|worker\_id\big)
    \end{flalign*}
\endgroup
where $\mathit{lrr_D}$, $\mathit{lrr_{BD}}$, and $\mathit{lrr_{CD}}$ were perceived causal support for a confounding, the gene effect on disease, and the gene effect on treatment, respectively, $\mathfrak{cs}_D$, $\mathfrak{cs}_\mathit{BD}$, and $\mathfrak{cs}_\mathit{CD}$ were our normative benchmarks corresponding to each log response ratio, $\mathit{delta\_p\_d}$ was the difference in the proportion of people with disease given gene versus no gene, $\mathit{delta\_p\_t}$ was the difference in the proportion of people with disease among those in the treatment group given gene versus no gene, $n$ was the sample size as a factor, $\mathit{vis}$ was a dummy variable for visualization condition, and $\mathit{worker\_id}$ was a unique identifier for each participant.

We primarily modeled effects on the mean of perceived causal support $\mu_D$, $\mathit{\mu_{BD}}$, and $\mathit{\mu_{CD}}$, but our models also learned residual standard deviations 
% of perceived causal support 
$\sigma_D$, $\mathit{\sigma_{BD}}$, and $\mathit{\sigma_{CD}}$.
The residual standard deviations and random effects in each model helped us separate patterns in responses from noise and individual differences.
In the first model, we used the term $\mathfrak{cs}_D*\mathit{delta\_p\_d}*\mathit{delta\_p\_t}*n*\mathit{vis}$ to learn how sensitivity to causal support for confounding varies as a function of sample size $n$ and visualization $\mathit{vis}$. 
In the second and third models, we used the terms $\mathfrak{cs}_\mathit{BD}*\mathit{delta\_p\_d}*n*\mathit{vis}$ and $\mathfrak{cs}_\mathit{CD}*\mathit{delta\_p\_t}*n*\mathit{vis}$ to learn how sensitive users in each visualization condition were to 
% the visual signal for 
the gene effects on disease $\mathit{delta\_p\_d}$ and treatment $\mathit{delta\_p\_t}$, respectively.

\noindent
\textbf{Qualitative analysis.}
We wondered how well participants would intuit how to perform the confounding detection task, considering it was more difficult than the task in Experiment 1, and we provided no training.
To address this we applied a deductive coding scheme.
We coded participants' strategy descriptions as \textit{uninformative} if they didn't describe a strategy. 
Otherwise, we coded whether or not participants described adequate strategies for judging \textit{delta p disease}, \textit{delta p treatment}, or \textit{sample size} (see Section 4.1.3), and we coded \textit{confusion} if they stated they were confused or described an incorrect strategy. 

\subsubsection{Participants \& exclusions}
% We recruited participants on Amazon Mechanical Turk using the same qualifications as in Experiment 1.
We used a similar approach to power analysis as in Experiment 1 to determine a target sample size of 500 participants after exclusions.
We recruited a total of 703 participants, and after exclusions we used data from 519 participants in our analysis. 
% Again, we slightly overshot our target sample size because we ran data collection in batches of participants, and we could not anticipate perfectly how many we would exclude. \yifan{Squeeze: prev sentence could go if short for space.} \jessica{agree}
Although we preregistered that we would exclude participants who allocated less than 25\% probability to confounding on an attention check trial where confounding was very likely (see Section 4.1.3), this criterion would have excluded 39\% of our sample.
We relaxed the cutoff to less than 20\% probability of confounding to allow for additional response error, resulting in a 26\% exclusion rate.
% \yifan{Exp 1's attention check was within 5\% of the target range?} This resulted in an exclusion rate of 26\% similar to Experiment 1.
% All participants were paid regardless of exclusions.
We paid participants \$3.04 for 14 minutes on average.
\subsection{Results}
We use a linear in log odds (LLO) model to describe performance in terms of \textit{sensitivity} to ground truth causal support and \textit{bias} in probability allocations when all four causal explanations are equally likely.

\noindent
\textbf{Sensitivity.} 
A LLO slope of one indicates ideal sensitivity to the log likelihood of the data given a set of causal explanations
% that users' probability allocations change in one-to-one correspondence with the log likelihood of the data given a set of causal explanations.
Similar to Experiment 1, slopes in all visualization conditions are closer to zero than one (Fig.~\ref{fig:e2-slopes}), indicating under-sensitivity to the ground truth.

Interacting with filterbars seems to improve sensitivity, while interacting with aggbars seems to decrease sensitivity, although these differences are not reliable.
% It is somewhat surprising to see a similar pattern of results for interactive visualizations in both experiments.
% We expected that interactive visualizations might be more helpful for detecting confounding  (Experiment 2) 
% than for detecting a treatment effect (Experiment 1).
It is surprising to see a similar pattern of results for interactive visualizations in both experiments, since we expected interactive visualizations to be more helpful for detecting confounding than for detecting a treatment effect.
% The confounding detection task requires users to look for more complex counterfactual patterns in order to distinguish between competing causal explanations, and the ability to manipulate data aggregation and filtering should enable users to query visualizations for these patterns.
Detecting confounding requires users to look for more complex counterfactual patterns in order to distinguish between causal explanations, and manipulating data aggregation and filtering should help users to query visualizations for these patterns.
When we analyze interaction logs (see Supplemental Materials), we see that filterbars users interacted with the visualizations more frequently and created more task-relevant views of the data than aggbars users, which may help to explain why interacting with filterbars was somewhat more helpful than interacting with aggbars.

\noindent
\textbf{Visual signal effects on sensitivity.}
We examine sensitivity in each visualization condition to the three visual signals for confounding in our task (delta p disease, delta p treatment, and sample size; see Section 4.1.3).
% For Experiment 2, the signal breaks down into three components: delta p disease, delta p treatment, and sample size (see Section 4.1.3). \yifan{not sure what the prev sentence is adding in this context. seems fine without?}
% \textit{Delta p disease} describes the difference in the proportion of people with disease in each data set depending on whether they have the gene.
% Negative values of delta p disease indicate evidence that the gene causes disease, whereas values near zero indicate evidence against a gene effect on disease.
% \textit{Delta p treatment} describes the difference in the proportion of people with disease within the treatment group depending on whether they have the gene.
% Negative values of delta p treatment indicate evidence that the gene stop the treatment from preventing disease, whereas values near zero indicate evidence against a gene effect on treatment.
% \textit{Sample size} is the number of people in each data set we showed chart users.
Normatively, slopes are one regardless of these visual signals.

Figure~\ref{fig:e2-interactions} shows users are more sensitive to causal support at values of delta p disease and delta p treatment near zero, with the exception of filterbars users who don't interact.
This pattern is consistent with the findings of Experiment 1 in that chart users respond more to evidence against a given causal effect than evidence in favor of an effect.

In Figure~\ref{fig:e2-interactions}, we also see that users of every visualization but filterbars are more sensitive when sample size is smaller. 
This pattern is consistent with prior work~\cite{Benjamin2016,Kim2019} and the results of Experiment 1.
%as well as prior work showing that people tend to underestimate sample size and underweight evidence from large samples in Bayesian update~\cite{Benjamin2016} including when making estimates from visualizations~\cite{Kim2019}.

\noindent
\textbf{Bias.}
LLO intercepts describe bias in probability allocations when the data are equally likely under each alternative causal explanation.
We derive expected probability allocated to explanation D based on LLO intercepts and compare this to the normative benchmark of 25\%.

Figure~\ref{fig:e2-intercepts} shows that, with all visualizations but text tables, users underestimate the probability of confounding in the absence of signal.
The fact that biases for each visualization condition differ between Experiments 1 and 2 suggests that these results are task-specific.
Future work should study reasons for these biases and what visual analytics software can do to help calibrate analysts' probability allocations. 
% \yifan{made a jump between biases and priors, prob just keep bias discussion or elaborate a bit more.}

\noindent
\textbf{Strategies.}
We assess users' self-reported strategy descriptions.
235 of 519 (45\%) users included in our analysis gave uninformative responses and were excluded from further analysis.
42 of 284 (15\%) remaining users either stated they were confused or described an incorrect strategy.

However, many users intuited the important signals in the data: 
\begin{quote}
    \vspace{-7pt}
    \textit{``I relied more on the `no treatment' cells to consider whether the gene causes the [disease], trying to look at ratio of `disease' and `no disease' within those two quadrants... [I] tried to consider the actual counts remembering that small numbers mean loose estimates but this was easy to overlook. Then I compared the two purple bars in the `gene no' top-half of graph to estimate the treatment effect... and did the same for the two lower purple bars to see if treatment equally effective in those with the gene.''}
\end{quote}
\vspace{-7pt}
\noindent
222 of 284 (79\%) described an adequate strategy for inferring the gene effect on disease.
81 of 284 (29\%) mentioned sample size information.
168 of 284 (59\%) described an adequate strategy for inferring the gene effect on treatment effectiveness.
These results suggest that much of our data represent a reasonable understanding of the task, yet participants still appeared to struggle to use the visualizations effectively.
\section{Discussion}
% We demonstrate the utility of causal support for studying inferences with visualizations.
% Our evaluation approach successfully measures some expected patterns in the quality of chart users' causal inferences.
We demonstrate the utility of causal support for evaluating inferences with visualizations, successfully measuring expected patterns in the quality of chart users' causal inferences.
For example, filterbars users should not have been able to perform either task without interacting because the visual signals required to perform the tasks were hidden behind interactions.
Our method shows that filterbars users were completely insensitive to the signal in data when they did not interact.
% with the visualizations.
Similarly, our models corroborate 
% findings from 
prior work suggesting that chart users underweight sample size when making inferences~\cite{Benjamin2016,Kim2019}.
Findings like these reassure us that causal support can help us understand how users struggle to use visualizations to evaluate causal hypotheses.

Our findings point to unsolved design challenges for supporting causal inferences with visual analytics (VA) tools.
Contrary to what we might expect given the emphasis of visualization research on evaluating encodings and interaction techniques, using different encodings for count data doesn’t appear to improve sensitivity to evidence for causal inferences beyond text contingency tables.
% Contrary to what we might expect given the emphasis on evaluating encodings and interaction techniques in visualization research, using different encodings for count data does not appear to improve sensitivity to evidence for casual inferences about possible data generating processes beyond text contingency tables.
Similarly, common interaction techniques in VA tools, such as manipulating data aggregation or cross-filtering coordinated multiple views, don’t seem to improve causal inferences beyond what users can achieve with simpler static visualizations.
Interacting with visualizations seems to help or hurt sensitivity depending on how deliberately signal-seeking users are and whether interacting is necessary in order to expose the visual signal in the data. 
This suggests that VA tools designed to optimize easy exposure of data are not sufficient for supporting causal inferences. 

We also find systematic biases in the way that chart users respond to specific visual signals in charts.
\textit{Chart users seem ubiquitously more sensitive to falsifying evidence than they are to verifying evidence}.
This may reflect a cognitive bias where analysts are more responsive to \textit{discrepancies}, between observed data and the counterfactual patterns expected under a given causal explanation, than they are to \textit{similarities} between observed data and counterfactual patterns.
Interestingly, this bias may be somewhat rational to the extent that verifying an inference is probabilistic, whereas the logic of falsification is deductive and thus ``more powerful'' in that it can definitively rule out an explanation~\cite{Popper1959}. 

\textit{Insensitivity to sample size remains a major challenge for informal statistical inferences}, and it appears not to be sufficiently addressed by common chart types for showing count data.
Even icon arrays, which emphasize sample size as the number of equal-sized dots, don't seem to mitigate this problem.
Prior work~\cite{Benjamin2016,Kim2019} suggests this may be due to perceptual underestimation of sample size and cognitive bias against claiming certainty in inferences.
Additionally, our qualitative results suggest chart users may not intuitively pay as much attention to sample size as they do to other signals when making causal inferences.
% As suggested by prior work~\cite{Benjamin2016,Kim2019}, this may be due to a combination of perceptual underestimation of sample size and cognitive bias against claiming certainty in inferences.
% Further, our qualitative results suggest that chart users may not intuitively pay much attention to sample size when making causal inferences. \alex{This breaks the flow too much}

Consistent with an aversion to believing causal relationships exist, we find that chart users tend to underestimate the probability of a given DAG arrow. 
In the absence of any signal differentiating between causal explanations, chart users allocate more probability to explanations that posit fewer 
% causal 
relationships, rather than allocating probability uniformly across alternatives.
Though this tendency interacts with task and visualization 
% format 
in ways that warrant further study, it may reflect an overall cognitive bias toward believing in simpler causal explanations. %with fewer assumptions.

\subsection{Limitations \& future work}
We set out to run a proof-of-concept study establishing causal support as an evaluation method for VA tools, and our study raises many unanswered questions.
A primary limitation of this work is that we recruited participants on Mechanical Turk, who may be less sensitive to causal support than real data analysts to the extent that they may use VA tools less deliberately.
However, our qualitative analysis suggests that many participants understood the task and used reasonable strategies.
Future work may find causal support helpful in evaluating current practices or novel interfaces with smaller pools of participants, insofar as real data analysts give less noisy responses than crowdworkers.
Questions remain about whether our findings generalize for other data types (e.g., continuous~\cite{Pacer2011} and event stream data~\cite{Pacer2015}), for domains outside of medicine, and for analysis scenarios with more complex possible data generating models.
Though we suspect our findings will persist in some form across user populations and analysis scenarios, visualizations probably will support some other causal inference tasks better than they support differentiating possible data generating processes.

% The primary limitation of this work is that we recruited participants on Mechanical Turk, who may differ from the user population for VA tools.
% Real data analysts may be more sensitive to causal support than crowdworkers to the extent that they may use VA tools more deliberately, though we suspect the overall patterns of biases that we find will persist in some form across user populations and causal inference tasks.
% We set out to run a proof-of-concept study establishing causal support as an evaluation method for VA tools.
% To the extent that real data analysts give less noisy responses than crowdworkers, the evaluation approach we demonstrate should be helpful in evaluating current practices or novel interfaces with smaller pools of participants.

\subsection{Improving visual analytics for causal inference}
A theme in visual causal inference is that analysts do not always know what to look for in data~\cite{Batanero1996,Yen2019}.
Causal inferences differentiating between possible data generating processes (DGPs) require comparisons between patterns in observed data and counterfactual patterns
% that should only manifest 
under a specific DGP~\cite{Greenland1999,Pearl2018}.
Users of 
% visual analytics 
VA software may 
% find causal inferences very challenging 
struggle with causal inferences
insofar as they fail to \textit{imagine} counterfactual predictions.
% from alternative DGPs.

Prior work in statistics and visualization argues for \textit{model checks} that make comparisons between data and model predictions explicit~\cite{gelman2003,gelman2004exploratory,hullman2020theories}.
For example, workflows in Bayesian statistics frequently employ prior and posterior predictive checks~\cite{Gabry2019}.%, where analysts pass draws from a prior or posterior distribution through a model in order to generate counterfactual predictions.
Visualizing model predictions alongside 
% the empirical distribution of the 
empirical data could support causal inference by externalizing discrepancies and similarities between observed and expected patterns.
% Such model checks may 
% improve performance on causal inference tasks
% support causal inferences insofar as they externalize the required comparisons, expressing information that is otherwise only in the imagination of the analyst.

We envision a 
% workflow for VA software 
VA workflow
where analysts cycle between interactively specifying models (e.g.,~\cite{Kraska2018}) and generating model checks to gauge model compatibility with their data. 
This echos calls to make models themselves a primary goal of visual data analysis~\cite{Andrienko2018}.
Causal support solves an important problem in realizing this vision, defining a ``good'' model check as one which supports sensitive inferences among a set of candidate 
% data generating models. 
DGPs.
Though it may be difficult to come up with an exhaustive set of 
% models for 
DGPs in
many real world applications, we think that this approach would be fruitful even with a relatively simple set of models that a knowledgeable analyst might provisionally entertain.
Causal support cannot guard against analysts ignoring possible models, but it can be used to evaluate visualization and interaction designs intended to help analysts collate and compare alternative models.

\section{Conclusion}
We contribute two crowdsourced experiments demonstrating an approach to evaluating causal inferences with visual analytics (VA) tools.
No visualization or interaction designs
% None of the visualization or interaction designs
we tested lead to reliably better causal inferences than text contingency tables, suggesting that common VA tools designed for data exposure may not be sufficient for supporting causal inferences.
We point to perceptual and cognitive biases which seem to make
visual
causal inferences difficult,
% for chart users, 
including tendencies to underweight both evidence verifying a causal relationship and evidence from large samples.
We discuss how formal models of causal support
% , such as the ones we demonstrate in our study, 
can be used to evaluate VA systems that place an emphasis on helping users reason about possible data generating processes.

\acknowledgments{
We thank the UW IDL and the NU MU Collective for their feedback.
We thank NSF (\#1930642) for funding this work.
}

\bibliographystyle{abbrv-doi}

\bibliography{0_causal-support}
\end{document}